\documentclass[11pt]{article}

\usepackage{amsmath,amssymb}
\usepackage{graphicx}
\usepackage{rotating}
\usepackage{fullpage}

\newcommand{\cP}{\mathcal{P}}

\newcommand{\cA}{\mathcal{A}}
\newcommand{\cB}{\mathcal{B}}
\newcommand{\cR}{\mathcal{R}}
\newcommand{\cV}{\mathcal{V}}

\newcommand{\N}{\mathbb{N}}
\newcommand{\Z}{\mathbb{Z}}
\newcommand{\R}{\mathbb{R}}
\newcommand{\tphi}{\tilde{\phi}}
\newcommand{\vtphi}{\boldsymbol{\tilde{\phi}}}
\newcommand{\vphi}{\boldsymbol{\phi}}
\newcommand{\vtpsi}{\boldsymbol{\tilde{\psi}}}

\newcommand{\epshat}{\hat{\varepsilon}}
\newcommand{\reset}{\rightarrow}
\newcommand{\p}{\varphi}
\newcommand{\s}{\sigma}
\newcommand{\phim}{\phi_{\textsf{m}}}
\newcommand{\eps}{\varepsilon}
\newcommand{\half}{\tfrac{1}{2}}
\newcommand{\Vone}{V_1}
\newcommand{\Vmax}{V_{\max}(\eps)}
\newcommand{\VMS}{V_{\textsf{MS}}}

\newtheorem{lemma}{Lemma}
\newtheorem{theorem}{Theorem}
\newtheorem{proposition}{Proposition}
\newtheorem{definition}{Definition}
\newtheorem{corollary}{Corollary}

\newcommand{\proof}{\noindent{\bf Proof:}~}
\newcommand{\qed}{\hfill $\square$ \newline \vspace{3mm}}
\begin{document}

\title{Unstable Attractors:  Existence and Robustness \\in Networks of
Oscillators 
With Delayed Pulse Coupling}

\author{Peter Ashwin$^1$\thanks{P.Ashwin@ex.ac.uk } \, and 
Marc Timme$^2$\thanks{timme@chaos.gwdg.de} \\
$^1$ Department of Mathematical Sciences, \\ Laver Building,
University of Exeter, Exeter EX4 4QE, UK\\
$^2$ Max-Planck-Institut f\"{u}r Dynamik und Selbstorganisation,\\
(formerly: MPI f\"{u}r Str\"{o}mungsforschung), \\ 
37073 G\"{o}ttingen, Germany.
}

\maketitle

\begin{abstract}
We consider unstable attractors; Milnor attractors $A$ such that, for
some neighbourhood $U$ of $A$, almost all initial conditions
leave $U$. Previous research
strongly suggests that unstable attractors
exist and even occur robustly (i.e.\ for open sets of parameter values) in
a system modelling biological phenomena, namely in globally coupled oscillators
with delayed pulse interactions.

In the first part of this paper we give a rigorous definition of unstable
attractors for general dynamical systems. We classify unstable
attractors into two types, depending on whether or not
there is a neighbourhood of the attractor
that intersects the basin in a set of positive measure. We give examples
of both types of unstable attractor; these examples have non-invertible
dynamics that collapse certain open sets onto stable manifolds of saddle
orbits.

In the second part we give the first rigorous demonstration of existence and
robust occurrence
of unstable attractors in a network of oscillators with delayed pulse coupling.
Although such systems are technically hybrid systems of delay
differential equations with discontinuous `firing' events, we show
that their dynamics reduces to a finite dimensional hybrid system
system after a finite time and hence we can discuss Milnor attractors for
this reduced finite dimensional system. We prove that for an open set of phase
resetting functions there are saddle periodic orbits that are unstable attractors.
\end{abstract}

\noindent {\bf Keywords}
Pulse-coupled oscillator, Heteroclinic cycle, Global coupling,
Hybrid system, Delay, Neural network.

\noindent {\bf PACS} 05.45.-a; 87.10.+e

\section{Background}

Attractors for dynamical systems are traditionally viewed as
being asymptotically stable invariant sets which have a neighbourhood 
that absorbs all sufficiently close initial conditions. As a consequence the 
basins of such attractors contain open sets. Whereas this concept works very 
well for simple attractors of ordinary differential equations, for more
complicated 
attractors the smallest asymptotically stable attractor
may contain much more than the asymptotic dynamics
of `typical' initial conditions.

Partly in response to this problem, Milnor introduced a
concept of measure attractor (now usually called {\em Milnor attractor})
\cite{Mil85}. This is a compact invariant set whose basins of attraction has 
positive measure in phase space. Examples of riddled basins \cite{Aleal92,Ott}
show that, even for smooth invertible dynamics, one can find Milnor
attractors with riddled basins; i.e. such that any open set that
intersects the basin has positive measure in the basin of
a different attractor; see \cite{AshParis04} for a recent review of riddled basins.

In the absence of smoothness or invertibility, it is clear that more exotic
attractors may appear. Recent work \cite{timmeprl,timmechaos,timmephd}
on globally coupled networks of oscillator with delayed pulse interactions
\cite{ernst95,ernst98} indicates that extreme cases of riddled basin attractors,
\emph{unstable attractors}, can appear, where there is a neighbourhood
$U$ of the attractor such that almost all points exit from $U$ under the
dynamics. Even more surprisingly, numerical simulations
\cite{timmeprl,timmechaos,timmephd} seem
to indicate that these attractors can appear in a robust way as long
as there are sufficiently many oscillators in the system. The main aim of this
paper is
to give a rigorous explanation for the appearance and robustness of
unstable attractors in such systems.

This paper is organized as follows. Section~\ref{secunstableatts}
examines the general problem of unstable attractors in a systematic
way. We give some
necessary conditions for the appearance of unstable attractors
and some motivating examples. Subsequently, we address
the relationship between networks of unstable attractors
and robust heteroclinic cycles, and present a simple recipe of how
to perturb a smooth flow with a robust heteroclinic cycle to
a smooth semiflow with a network of unstable attractors.

In Section~\ref{secpulsecos} we consider the system of globally pulse
coupled oscillators with delayed interactions studied in \cite{timmeprl},
including a generalization of
the set of possible response functions. Because the system
combines continuous time evolution with discrete time events,
it is strictly speaking a hybrid system \cite{hybrid} with delay.
The main result of this section
is Theorem~\ref{thmfinite} which shows that the dynamics reduces,
after a finite time, to a finite dimensional hybrid system. It gives
an explicit upper bound on the number of remaining dimensions in the system.

In Section~\ref{secuasforpulsecos} we prove
that the system discussed in Section~\ref{secpulsecos} exhibits unstable
attractors (Theorem~\ref{thm:uasexist})
and that these persist for an open set of parameter values
(Theorem~\ref{thm:uasrobust}).
The proof has two parts; firstly we show (by following an appropriate open set
of initial conditions) that a certain periodic orbit is a Milnor attractor;
secondly we show that this orbit is a saddle. We discuss this mechanism as
being an effect of `dimension jump' in the system.

Section~\ref{secdiscuss} discusses
obstacles to proving the existence and robustness of unstable
attractors in a more general setting.
For convenience some of the details of the proofs have been placed in
appendices: Appendix~\ref{App:k=1} describes the system in
the case that the delay is short enough that only one delayed pulse can
influence the future dynamics of the system.
Appendix~\ref{App:unstable} derives the return map on a section transverse
to the unstable attracting periodic orbit.

\section{Unstable attractors}
\label{secunstableatts}

\subsection{Milnor attractors and unstable attractors}

Consider a dynamical system defined by a semiflow $F_{t}:M\rightarrow M$
on a finite dimensional manifold $M$ for $t \geq 0$ (recall that
a semiflow is a family of maps such that $F_{s+t}= F_{s} \circ F_t$
whenever $s\geq 0$, $t\geq 0$, and such that $F_0(x)=x$). In what follows
we will consider continuous ($t\in \R^+$) or discrete ($t\in\Z^+$) dynamics.

Our examples arise from cases where $F_t$ is not invertible for all $t>0$:
one cannot extend the semiflow to a flow.
For convenience we will assume that $M$ is compact and let $\ell(\cdot)$ denote
Lebesgue/Riemann measure on $M$. Recall that the $\omega$-limit set of $x$ is
defined by
\begin{equation}
\omega(x)=\bigcap_{t>0}\overline{\{ F_{s}(x)~:~s\geq
t\}}.\label{eq:OmegaLimitSet}
\end{equation}
A {\em Milnor attractor} for $F_{t}$ is a compact invariant subset $A\subset M$
such
that (i) the basin of attraction
\begin{equation}
\cB(A)=\{ x\in M~:~\omega(x)\subset A\}\label{eq:basin}
\end{equation}
has positive measure in $M$, i.e. $\ell(\cB(A))>0$, and (ii) any
compact invariant proper subset of $A$ has a basin with strictly
smaller measure \cite{Mil85}. It is a {\em minimal Milnor attractor}
if any compact invariant proper subset of $A$ has a basin with zero measure. All
explicit examples of Milnor attractors given in this paper are fixed points or
periodic orbits and thus they are minimal Milnor attractors because
they do not contain any compact invariant proper subset. However, we will not
explicitly discuss minimality in these cases.

\begin{definition}[Lingering set]
Given any subset $U\subset M$ we define the {\em lingering subset} of $U$ to be
\begin{equation}
\cA(U)=\{x\in U~:~ F_t(x)\in U \ \mbox{ for all } t\geq 0 \}.
\end{equation}
\label{def:lingeringset}
\end{definition}
The lingering set is a (forward) invariant set consisting of
points in $U$ which do not leave $U$ in the future;
note that if $A$ is an invariant set within $U$ then $\cA(U)$ is non-empty as it contains $A$.

\begin{definition}[Unstable Attractor]
\label{def:UA}
We say a Milnor attractor $A$ is an \emph{unstable attractor} if there is a
neighbourhood
$U$ of $A$ with
\begin{equation}
\ell(\cA(U))=0.
\label{eq:NoContractingBasin}
\end{equation}
\end{definition}
Requirement (\ref{eq:NoContractingBasin}) means that almost all
trajectories in a neighbourhood of the attractor must leave this
neighbourhood. We distinguish between two classes of unstable attractors.

\begin{definition}[\bf Unstable Attractor With a Positive (Measure) Local Basin]
\label{def:UApositivelocalbasin}
An unstable attractor $A$ is an {\em unstable attractor with
positive measure local basin}\footnote{For convenience of
notation we refer to {\em positive local basin} (resp. {\em zero local 
basin}) for Def.~\ref{def:UApositivelocalbasin} (resp. 
Def.~\ref{def:UAzerolocalbasin}).} if
\begin{equation}
\ell(\cB(A)\cap U)>0\label{eq:UAlocalbasin}
\end{equation}
for all neighbourhoods $U$ of $A$.
\end{definition}

\begin{definition}[\bf Unstable Attractor With a Zero (Measure) Local Basin]
\label{def:UAzerolocalbasin}
An unstable attractor $A$ is an \emph{unstable attractor with zero measure local
basin}
if there is a neighbourhood $U$ of $A$ with
\begin{equation}
\ell(\cB(A)\cap U)=0.  \label{eq:UAnoLocalBasin}
\end{equation}
\end{definition}

For an unstable attractor with positive local basin,
(\ref{eq:NoContractingBasin}) means that almost all trajectories starting in a
sufficiently small neighborhood $U$ 
will first leave $U$. A positive measure will eventually return and be
asymptotic to $A$ (contrasting with with \cite{Mendelson:270:1960,Melbourne:835:1991} 
where there is a positive measure that remains in $U$ and is asymptotic to $A$).

For an unstable attractor $A$ with zero local basin almost
all initial conditions in a sufficiently small neighbourhood $U$ of
$A$ will leave $U$ and never limit to $A$.\footnote{
A set $B$ has positive measure if $\ell(B)>0$; it has full measure (in $U$) if 
$\ell(B\cap U)>0$ and $\ell(U\setminus B)=0$.
}

Clearly, linearly stable fixed points or periodic orbit attractors
for a smooth invertible system cannot be unstable attractors;
in such cases one can construct a neighbourhood that by the Hartman-Grobman
Theorem
is contained within the basin of attraction.
If $A$ is asymptotically stable then trivially it
cannot be an unstable attractor. A more general observation is the following
proposition characterizing a class of attractors that
cannot be unstable with zero local basin.

\begin{proposition}
Suppose that \(A\) is a Milnor attractor whose basin \(\cB(A)\) contains an open set,
and suppose that \(F_{t}\) is almost everywhere a local homeomorphism.
Then \(A\) cannot be an unstable attractor with zero local basin.
\label{prop:nounsts}
\end{proposition}

\proof
Consider a point $y$ in the interior of the basin of $A$ such that
$F_t$ is a local homeomorphism in a neighbourhood of $y$, and choose
an open neighbourhood $U$ of $A$. There is a finite time $T$
after which $F_{T}(y)\in U$. Pick a small neighbourhood $V$ of $F_{T}(y)$
in $U$; continuity of $F_{T}$ means that there is a neighbourhood $W$
of $y$ within the basin of attraction such that $F_{T}(W)\subset V$.
Continuity of the local inverse means that $F_{T}(W)$ is also a neighbourhood
of $F_{T}(y)$. However all points in $W$ are in the basin of $A$ and so
$F_{T}(W)$ is an open set in the basin of $A$ that is within $U$. Hence 
$\ell(\cB(A)\cap U)>0$ for any $U$ that is a neighbourhood of $A$. 
\qed

In particular, this proposition ensures that continuous invertible flows with
continuous inverse (homeomorphisms) do not exhibit unstable attractors with
zero local basin. We have so far been unable to determine whether the
assumptions of Proposition~\ref{prop:nounsts} exclude the possibility that
$A$ is an unstable attractor with positive measure local basin, though clearly
under additional assumptions (such as linear stability of $A$) we can exclude
this case.

As we will see below, unstable attractors occur robustly
for certain classes of noninvertible dynamical systems.
Suppose $X$ is a space of dynamical systems under some suitable topology, such
that one can obtain an unstable attractor for some $f\in X$. If there are
unstable attractors for all $g$ in a neighbourhood of $f$ then we say the unstable
attractors for $f$ are \emph{robust}; this means that they are stable to perturbations of
the system within this $X$. In Section~\ref{secuasforpulsecos}
we show robustness of unstable attractors with respect to perturbations
of a pulse response function defining a class of pulse coupled oscillator
systems.

\subsection{Unstable attractors and riddled basin attractors}

There is no simple relationship between the lingering set for a given
neighbourhood of an attractor and the basin of attraction; in particular there may 
be points that are in the lingering set and not in the basin, and vice-versa. For example,
near a centre in a planar system the lingering set can be an open set but the basin just
one point. Similarly the flow of $\dot{x}=1-\cos(2 \pi x)$ on the unit circle $\R/\Z$
has a fixed point $x=0$.  The basin of the attractor $A={0}$ is the whole of phase space, 
but the lingering sets of small open neighbourhoods of $x=0$ will not contain nearby
points to the right of $x=0$.

Unstable attractors are a related but different phenomena to riddled basin attractors $A$ \cite{Aleal92}.
Recall that an attractor $A$ has \emph{riddled basin} if $\cB(A)$ is such that
\begin{equation}
\ell(\cB(A)\cap U) \ell (\cB(A)^c \cap U) >0
\end{equation}
for all open sets $U$ that intersect $\cB(A)$. In particular, any neighbourhood
$U$ that contains $A$ will intersect $\cB(A)$ in a set of positive measure; this means
that a positive measure set in the basin leaves $U$. For unstable
attractors almost all points in a small enough $U$ will leave $U$; by the comment above,
this does not have any direct implication for the basin of attraction of $A$, though
it does imply that any `local basin of attraction' relative to $U$ will have zero
measure. Examples of riddled basin attractors so far all possess irregular or chaotic
dynamics to provide attraction local to some points and repulsion local to others. 
By contrast the dynamics observed on unstable attractors can be very simple, even 
equilibrium.

\subsection{Simple examples of unstable attractors}

\begin{figure}
\begin{center}\includegraphics[width=8cm]{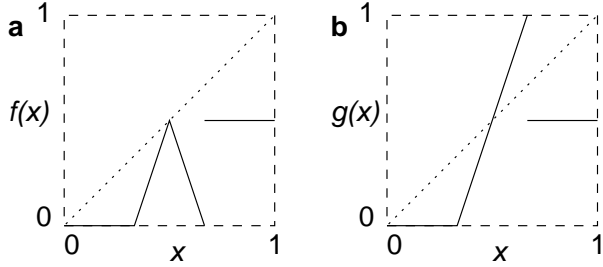}
\end{center}
\caption{ 
Dynamical systems defined by piecewise smooth maps 
$f$ and $g$ exhibit an unstable attractor 
$A=\half$ with (a) zero and (b) positive local basin. 
\label{figeg1}}
\end{figure}

To illustrate the possible types of unstable attractors we consider
the dynamical systems defined by the piecewise continuous mapping
$f:[0,1]\mapsto[0,1]$ according to
\begin{equation}
f(x)=\left\{ \begin{array}{cl}
0 & \mbox{ if }0\leq x<\tfrac{1}{3}\\
3(x-\tfrac{1}{3}) & \mbox{ if }\tfrac{1}{3}\leq x<\half\\
-3(x-\tfrac{2}{3}) & \mbox{ if }\half\leq x<\tfrac{2}{3}\\
\tfrac{1}{2} & \mbox{ if }\tfrac{2}{3}\leq
x\leq1\end{array}\right.\label{eqeg1}
\end{equation}
and $g:[0,1]\rightarrow[0,1]$ where $g(x)=f(x)$ except on $\half\leq
x<\frac{2}{3}$
in which case $g(x)=3(x-\frac{1}{3})$. These two maps are illustrated in
Figure~\ref{figeg1}. One can
verify that the only invariant sets for $f$ and $g$ are the two fixed points
$x=0$ and $x=\frac{1}{2}$. These are both minimal Milnor attractors; for $f$
their basins of attraction are respectively
\begin{equation}
\cB(\{0\})={\textstyle
[0,\tfrac{1}{2})\cup(\tfrac{1}{2},\tfrac{2}{3}),~~\cB(\{\tfrac{1}{2}\})
=\{\tfrac{1}{2}\}\cup[\tfrac{2}{3},1]}.\label{eq:basinf(0)}
\end{equation}
whereas for $g$ they are
\begin{equation}
\cB(\{0\})=[0,\half),~~\cB(\{\half\})=[\half,1].\label{eq:basing(0)}
\end{equation}
In both cases the disjoint union of the two basins is the interval $[0,1]$.
On the other hand, any open set $U\subset (\tfrac{1}{3},\tfrac{2}{3})$ that contains
$\tfrac{1}{2}$ has lingering set $\cA(U)$ that is just the point $\{\tfrac{1}{2}\}$. Hence
in both cases this point is an unstable attractor. Considering the definitions 
(\ref{eq:UAlocalbasin}) and (\ref{eq:UAnoLocalBasin}) above, we see that the attractor $A=\{\half\}$ 
is an unstable attractor
with zero local basin for $f$ and an unstable attractor with positive local
basin for $g$.

\subsection{Networks of unstable attractors}

As noted in \cite{timmechaos}, unstable attractors can form networks
such that a positive measure of sufficiently small perturbations from the
unstable 
attractor $A_{i}$ are in the basin of attraction of some other unstable 
attractor $A_{j}$. We show that this is analogous to robust heteroclinic
networks.

\begin{definition}
A set of unstable attractors $A_{i}$, $i\in \{1,\cdots,n\}$ forms a
\emph{network} given by
the directed graph with vertices $A_{i}$. There are directed edges from $A_{i}$
to $A_{j}$
if and only if for any neighbourhood $U_{i}$ of $A_{i}$ we have
\begin{equation}\label{eq:networkofuas}
\ell(\cB(A_{j})\cap U_{i} )>0.
\end{equation}
\end{definition}

\begin{figure}
\begin{center}
\includegraphics[width=4cm]{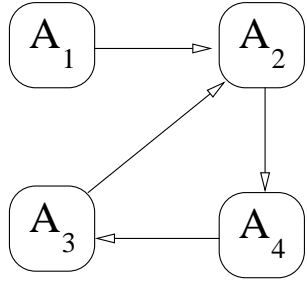}
\end{center}
\caption{
A closed network of unstable attractors. The unstable attractors $A_{i}$ may
form a  network where a link from $A_i$ to $A_j$ implies that the basin $\cB(A_j)$
intersects every neighbourhood of $A_i$ in a set of positive measure. In the example
sketched above, there is a transitive
subnetwork $\{ A_{2},A_{3},A_{4}\}$. \label{fignetworkofuas}
}
\end{figure}

If in addition
\begin{equation}
U_i = \bigcup_{k} \cB(A_k)\cap U_i.
\end{equation}
we say the network is \emph{closed}. All trajectories near a closed network
cannot leave a neighbourhood of the network.
In such a case the dynamics of the system is well described by
the network even in the presence of small perturbations to the system. We
illustrate such
a network schematically in Figure~\ref{fignetworkofuas}.

If a network is not closed, we only have
\begin{equation}\label{eq:notclosed}
\ell(U_i) \geq \sum_{j} \ell(\cB(A_j)\cap U_i)
\end{equation}
and such a network of unstable attractors $A_i$ may still manifest itself in 
transients. If there is a proper inequality in (\ref{eq:notclosed}) then a
positive measure of initial conditions starting near the network are not asymptotic 
to states that are in the network.

A network is called \emph{transitive} if there is a
directed path between any two attractors. An arbitrary closed network will 
contain subnetworks that are transitive networks of unstable attractors.

\subsection{Unstable attractors from saddles}
\label{sechetcycles}

We briefly demonstrate that one can obtain unstable attractors
from smooth semiflows with saddles by including small
perturbations that induce collapse onto their stable manifolds.

Consider a hyperbolic saddle equilibrium $x_{0}\in M$ for a (non-invertible)
semiflow $F_{t}:M\rightarrow M$ on a compact $M$ with stable manifold
$W^{s}(x_{0})$. We say $F_{t}$ exhibits \emph{collapse onto the
stable manifold} of $x_{0}$ if there is an open set $U\in M$ and
a $T>0$ such that $F_{T}(U)\subset W^{s}(x_{0})$. Since
$x_0$ is a saddle, and has thus an unstable manifold, such an $U$ cannot contain
$x_0$.
Similarly, one can define \emph{collapse onto the
unstable manifold} of a saddle $x_0$. This can occur on
a neighbourhood that includes $x_0$ and in this case $U$ may contain
$x_0$.

\begin{figure}
\begin{center}
\includegraphics[width=80mm]{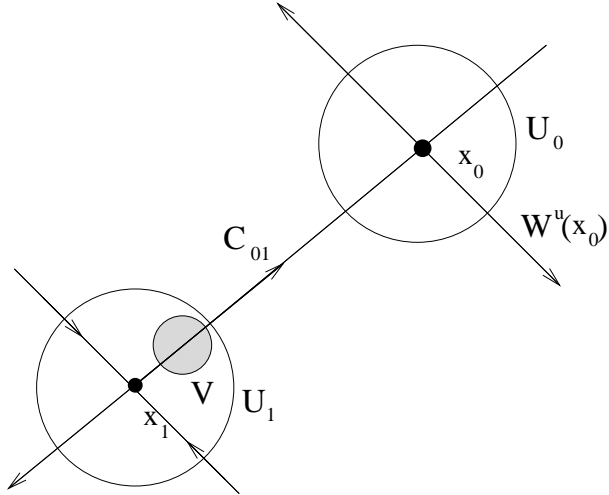}
\end{center}
\caption{
A semiflow on the plane with two saddles; $x_{0}$ is an unstable 
attractor. The set $C_{01}=W^u(x_1)\cap W^s(x_0)$ is a heteroclinic connection
from $x_1$ to $x_0$. We assume that the open set $V$ in the neighbourhood $U_{1}$ 
of another equilibrium $x_1$ collapses in finite time onto
the stable manifold $W^s(x_0)=\cB(x_0)$ of $x_{0}$. Hence $\ell(\cB({x_0}))>0$; moreover, 
almost all perturbations to $x_0$ within $U_0$ will leave $U_0$ and so
$x_0$ is an unstable attractor.
\label{fig:twosaddles}
}
\end{figure}

An example of an unstable attractor constructed by
using a pair of saddle equilibria is shown in
Figure~\ref{fig:twosaddles}; this is essentially the mechanism that
causes the appearance of unstable attractors in the system we discuss
in Section~\ref{secpulsecos}.
If there is a connecting orbit from $x_1$ to $x_0$ then
collapse onto the unstable manifold of $x_1$ is simultaneously a
collapse onto
the stable manifold of $x_0$; the points in the open set $V$ intersecting
the unstable manifold of $x_{1}$ tend to $x_{0}$.
Unstable attractors in this sort of scheme can be robust to perturbations of the
system if the connecting orbits are in invariant subspaces.

\subsection{Networks of unstable attractors and heteroclinic networks}

Smooth dynamical systems with symmetries can exhibit heteroclinic cycles
or homoclinic cycles as robust Milnor attractors,
if the connecting orbits lie within symmetry-forced invariant subspaces; see for
example \cite{Kru97,KM}. We indicate briefly how one can use these to construct
a large class of systems
that have networks of unstable attractors.

Consider a smooth semiflow $F_{t}:M\rightarrow M$ on $M$ with an
absorbing open region $R\in\R^{n}$ such that the only Milnor attractor
in $R$ consists of the set
\begin{equation}
\Sigma=\bigcup_{i}\{ x_{i}\}\cup W^{u}(\{ x_{i}\}).\label{eq:hetcycle}
\end{equation}
For hyperbolic saddle equilibria $x_{i}$ and the unstable manifolds 
$W^u(x_i)$. We assume that the unstable
manifolds are one dimensional and write the connecting orbits from
$x_{i}$ to $x_{j}$ by
\begin{equation}
C_{ji}=W^{u}(x_{i})\cap W^{s}(x_{j}).\label{eq:hetconnection}
\end{equation}
Given such an attractor $\Sigma$, we can construct piecewise
smooth semiflows $\tilde{F}_{t}:M\rightarrow M$
such that the set of $x_{i}$ forms a network of unstable attractors,
and such that $F_{t}$ and $\tilde{F}_{t}$ are identical except on
a set of arbitrarily small measure.

\begin{figure}
\begin{center}
\includegraphics[width=8cm]{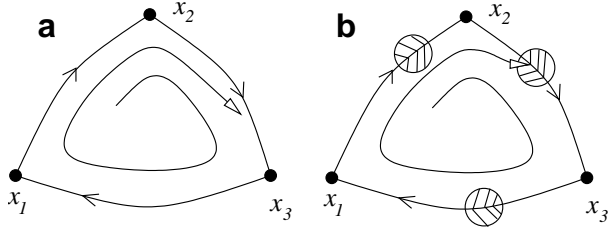}
\end{center}

\caption{
From heteroclinic networks to networks of unstable attractors. (a) An attracting
heteroclinic cycle between three hyperbolic equilibria $x_i$ can be locally
perturbed in sets of arbitrarily small measure to obtain (b) a network of unstable
attractors with 
the same connection topology. Here, trajectories entering the balls hit one of
the connecting orbits after a finite time and continue out after a uniformly bounded
time.
\label{figuasfromhets}
}
\end{figure}

To do this, pick a section $S_{ij}$ to each connection $C_{ji}\neq\emptyset$
and a neighbourhood $U_{ij}$ of $S_{ij}$ of arbitrarily small measure.
Define $\tilde{F}_{t}$ to be a
semiflow that is equal to $F_{t}$ when the trajectories are outside
of the $U_{ij}$. On entering the $U_{i}$ we require that the semiflow collapses
onto the connecting orbit in finite time, as illustrated in
Figure~\ref{figuasfromhets}.
The illustration depicts this for one-dimensional connections but
there can clearly be higher-dimensional connections with qualitatively the same features.

Note that robust networks of unstable attractors can appear under the same
conditions as for robust heteroclinic attractors \cite{Kru97} by
adding collapse onto {\em unstable} manifolds because in this case unstable
manifolds are
contained within stable manifolds of other saddles. This can arise in
cases where there is a discontinuity in the semiflow, but we note
that it is not necessary; one could construct an unstable attractor
similarly to Figure~\ref{fig:twosaddles} for a smooth semiflow,
as long as this maps some open set onto a zero measure subset of the
stable manifold of $x_{1}$.

\section{Oscillators with delayed pulse-coupling:
\newline a rigorous reduction to a finite dimensional system}
\label{secpulsecos}

In this section we consider the system of oscillators with delayed pulse
interactions
investigated in \cite{timmeprl}. First we will show that this system, although
formally infinite dimensional, can be reduced under general
assumptions to a finite dimensional system. The number of dimensions
remaining depends on the system parameters.

\subsection{From phase history functions to phases and firing times}

The system of oscillators we consider is technically a hybrid dynamical
system. A hybrid dynamical system is defined by a smooth flow in some regions 
of phase space with boundaries; when the flow hits one of the boundaries it is then
mapped instantaneously by a deterministic mapping to some other point on the boundary.
A trajectory from a hybrid dynamical system hence consists of intervals of
continuous evolution interrupted by events that may be discontinuous.
An evolution of the system is a {\em Zeno State} \cite{Zeno}
(also referred to as a {\em chattering state})
if there can be an infinite number of applications of the map in
some finite time interval; in such cases one needs to have extra rules
to permit evolution beyond the first accumulation point in time of
an infinite number of applications of the map. We will show that there are no
Zeno 
states in the system considered below.

\begin{definition}[Phase History Functions]
\label{def:phasehistory}
We define $\cP$ to be the set of piecewise smooth
`phase history' functions
$\tphi:(-\infty,0]\rightarrow[0,1)$, $s\mapsto\tphi(s)$ such that:
\begin{itemize}
\item[(i)] {\em Well defined limit from below.} For some discrete (possibly
finite), monotone 
increasing set of
$s_{k}\geq 0$, $k\in\mathbb{{N}}$ we have $\tphi(s)$ is smooth for  $s\in
[-s_{k+1},-s_{k})$.
\item[(ii)] {\em No Zeno states.} If there are infinitely many discontinuities
$s_k$ then $\lim_{k\rightarrow\infty}s_{k}=\infty$.
\end{itemize}
\end{definition}
Note that (i) means that the limit from below
$\tphi(s'-):=\lim_{s\rightarrow s'}\tphi(s)$ is defined for all $s'\leq 0$.
Condition
(ii) means that we allow no Zeno states in the past.
The system \cite{timmeprl} of $N$ pulse-coupled oscillators with delay $\tau$,
where the phase histories are $\vtphi=(\tphi_{1},\cdots,\tphi_{N})$,
is defined in terms of a semiflow on $\cP^{N} \ni \vtphi$ for all $t\geq0$
\begin{equation}
\Phi_{t}:\cP^{N}\rightarrow\cP^{N}\label{eq:semiflowforoscs}
\end{equation}
such that $\Phi_{t}(\vtphi)(s)=\vtphi(s+t)$ for all $s\leq -t$.
On the interval $s\in(-t,0]$ the function $\Phi_{t}(\vtphi)(s)$
is defined below. The boundaries of the intervals on which
the phase history function
$[\Phi_t(\vtphi)]_i$ of oscillator $i$ is smooth are
now denoted by $-s_{i,k}(t)<0$, $k\in \N$.

If $\lim_{s\rightarrow s'- }\tphi_{i}(s)\geq 1$ we say oscillator $i$ 
\emph{fires}
at time $s'$. Given that the system has evolved a time $t$ from some initial
condition $\vtphi\in\cP^{N}$ at time $0$, we list the past firings
$-\s_{i,k}\leq 0$ of
oscillator $i$ as
\begin{equation}
\cdots<-\s_{i,3}(t)<-\s_{i,2}(t)<-\s_{i,1}(t)\leq 0
\label{eq:firings}
\end{equation}
such that the  $\s_{i,k}(t)$ are the times since the
$j$th last firing of the $i$th oscillator before or at time $t$. Note that the
$\s_{i,k}$, 
$k\in \mathbb{N}$ count the firing events 
whereas the $s_{i,k}$,
$k\in \mathbb{N}$ count all discontinuities that may occur. We
write the phase of the oscillator $i$ as $\phi_{i}=\tphi_{i}(0)$,
namely as the most recent point in the phase history of that oscillator,
and the previous firing times as $\s_{i,k}$.

The future dynamics will be determined only by the present phases and the
times of the past firings; formally there is a map
$\Pi:\cP^N \rightarrow (\R^{+})^{\N}$ such that
\begin{equation}
\Pi(\vtphi)=\left(
\phi_{i}\,,\left(\s_{i,k}\right)_{k\in\mathbb{{N}}}
\right)_{i\in\{1,\ldots,N\}}.
\end{equation}

Thus the map $\Pi\circ \Phi_t(\tphi)(0)$ directly gives the current phases
$\phi_i(t)$ and the past firing times $\s_{i,k}(t)$,
$k\in\mathbb{{N}}$, of the oscillators $i\in\{1,\ldots,N\}$.
Similarly, it will be useful to define a map onto the phases
and the $k$ past firing times:
\begin{equation}
\Pi_{k}(\vtphi)=\left(
\phi_{i}\,,\left(\s_{i,k'}\right)_{k'\in\{1,\ldots,k\}}
\right)_{i\in\{1,\ldots,N\}}.
\end{equation}

\subsection{Definition of the semiflow $\Phi_{t}$}\label{sect:defsemiflow}

The semiflow $\Phi_{t}$ in (\ref{eq:semiflowforoscs}) is given as
a hybrid system with continuous time evolution
\begin{equation}
\label{eq:phisdotone}
\frac{d \phi_i}{d t}=1,~~\frac{d \s_{i,k}}{d t}=1
\end{equation}
for all $i$ and $k$, except when one of a number of possible discrete events occur.

These events are determined by a constant delay $\tau>0$ between one oscillator being 
reset and `sending' a pulse (`firing') and its `reception' by another oscillator. 
These events `firing' and `reception' 
are the only types of discrete events interrupting the continuous time evolution. 
We say that at a time $T$
\begin{itemize}
\item
{\em oscillator $i$ fires}, \\ denoted $S_i$, if $\phi_{i}(T-)=1$.
\item
{\em oscillator $i$ receives a pulse from oscillator $j$}, \\
denoted $R_{ij}$, if $\s_{j,k}(T-)=\tau$  for some $k$ and $\phi_{i}(T-)<1$.
\item
{\em oscillator $i$ fires induced by reception of a pulse from oscillator
$j$}, \\
denoted $S'_{i}$, if $R_{ij}$ and this results in
an immediate firing of oscillator $i$ (see below).
\end{itemize}

Observe that for any $i$ it is not possible that $S_{i}$ and $R_{ij}$
(for any $j$) occur simultaneously, whereas if $R_{ij}$ occurs it
is possible that $R_{ij'}$, $j'\neq j$ also occurs at the same time.
The phases
$\vphi(t)=(\phi_i(t))_{i\in\{1,\ldots,N\}}$
and firing times
$\sigma(t)=(\sigma_{i,k})_{i\in\{1,\ldots,N\},k\in\N }$, and thus the evolution
$\Phi_t(\vtphi)$ occurs as (\ref{eq:phisdotone}) until one of $S_{i}$, $R_{i,j}$ or $S'_i$ occur 
at time $T\geq t$. By Definition~\ref{def:phasehistory}, we take 
$\phi_i(t)\in \left[0,1\right)$ and $\s_{i,k}(t)\geq 0$ for all $i,k$ and all
$t$.  At such an event we define the new phases using the list below.
One can verify that all variables stay in their respective ranges if 
they are started in that range and evolving in this way.

\begin{itemize}
\item {\em For each $i$ such that $S_{i}$ occurs} \\
we set
\begin{equation}
\phi_{i}(T):=0, \ \s_{i,1}(T):=0 \mbox{ and renumber }
\s_{i,k}(T):=\s_{i,k-1}(T-) \mbox { for all } k\geq 2.
\end{equation}
\item {\em For each $i$ such that $R_{ij}$ occurs} \\
we set
\begin{equation}
\phi_{i}(T):=\phi_{i}(T-)+V(\phi_{i}(T-),\epshat_{i})
\label{eq:epshat}
\end{equation}
where $V(\phi,\eps)$ is defined below,  $\epshat_{i}=\frac{\eps K_{i}}{N-1}$,
$\eps>0$, and
\begin{equation}
K_{i}=\#\{ l\in\{1,\ldots,N\}~:~\s_{l,m}(T-)=\tau\mbox{ for $l\neq i$ and some
$m$}\}
\label{eq:kinputsnow}
\end{equation}
is the number of `input pulses' that arrive at time $t=0$.
\item Once all the $R_{ij}$ have been checked, we say there is
{\em firing $S'_i$ induced by pulse reception} if $R_{ij}$ occurs for
some $j\neq i$ and results in $\phi_{i}(T)\geq1$. In this case
we set $\phi_{i}(T):=0$,\ $\s_{i,1}(T)=0$ and
renumber $\s_{i,k}(T)=\s_{i,k-1}(T-)$ for all $k\geq 2$, as if $S_i$ occurred.
\end{itemize}
One can verify that the order of `processing' simultaneous
events with different $i$
does not affect the final outcome; hence the map is well-defined.
The condition $l\neq i$ in (\ref{eq:kinputsnow}) excludes self-interaction
as in \cite{timmeprl,timmechaos,timmephd} which we will consider
exclusively in the following.
The response function $V(\phi,\epshat)$ expresses the response of an oscillator
at phase
$\phi$ to an input $\epshat$.

The model is specified once the delay $\tau > 0$, the coupling strength $\eps >
0$
and the function $V$ are known.
Observe that given an initial $\vtphi$ and $\phi(t)$ for $0<t\leq T$
we can define $\Phi_{T}(\vtphi)$ the phase history at time $T>0$.
This way we complete the definition of  $\vphi_t(\tphi)(s)$
(\ref{eq:semiflowforoscs})
for times $s\in(-t,0]$.

For continuous functions $V:[0,1]\rightarrow \R$ we define their {\em difference
increments}

\begin{equation}
\label{DeltaV}
\Delta_V(\phi,\eps):= \min_{0\leq \phi' \leq 1-\phi}
\left(V(\phi'+\phi,\eps)-V(\phi',\eps)\right).
\end{equation}
We say $V$ is {\em uniformly strictly monotonic increasing} on $[0,1]$ if 
$\Delta_V(\phi,\eps)>0$ for all
$0 < \phi\leq 1$ (note that this implies that $V(\phi,\eps)$ is
strictly monotonic increasing in $\phi$).

\begin{definition}
[Response Functions with Increasing Response]
\label{def:cV}
We define $\cV$ be the set of response functions
$V: [0,1]\times [0,c] \rightarrow [0,\infty)$ for some $c>0$
with $V(\phi, \eps)$ satisfying
\begin{itemize}
\item[(i)]
{\em Continuity: } $V(\phi,\eps)$ is jointly continuous in $\phi$ and $\eps$.
\item[(ii)]
{\em Zero response to zero input: } $V(\phi,0)=0$ for all $\phi$.
\item[(iii)]
{\em Uniform strict monotonicity:}
For any fixed $\eps>0$ the function $V(\phi,\eps)>0$ is
uniformly strictly monotonic increasing in $\phi$ on $[0,1]$.
\item[(iv)]

{\em Upper bound linear in $\eps$:}
For any fixed $\eps>0$, monotonicity of $V$ means there is a unique $\phim$ that
solves
\begin{equation}
\label{eq:phim}
V(\phim,\eps)+\phim=1
\end{equation}
which gives the maximum response; we assume this is bounded above by a linear
function of
$\eps$. More precisely we assume there is a $V_1$ such that
\begin{equation}
\label{eq:Vone}
V(\phim,\eps) \leq V_1 \eps \leq 1
\end{equation}
for all $0\leq \eps \leq c$.
\end{itemize}
\end{definition}
Condition (iv) allows one to obtain a bound on the total change in phase over
one period.
Note that $V(\phi,\eps)\leq\Vone\eps<1$ for all $\phi\leq \phim$ and that 
for $\phi>\phim$, the response $V(\phi,\eps)$ is such that the updated 
phase (\ref{eq:epshat}) is supra-threshold, resulting in an induced firing;
hence $\Vmax=V(\phim,\eps)$
is the largest response to an input of magnitude $\eps$.

We say that response functions $V\in\cV$ give {\em increasing response} because
the 
response increases

with $\phi$. In particular, concave (down) potential functions give rise to
this form of response (see Lemma~\ref{lemma:UV} below).
We give this space the structure of a metric space with norm
\begin{equation}
\label{eq:Vnorm}
\|V- \tilde{V} \| = \max_{(\phi,\eps)\in[0,1]\times[0,c]}
|V(\phi,\eps)-\tilde{V}(\phi,\eps)|
\end{equation}
for $V,\tilde{V}\in \cV$. Closeness of $V$ and $\tilde{V}$ in this norm imply
that both the
functions and their differences increments are close, by the following Lemma.

\begin{lemma}
\label{lemma:Deltabound}
Suppose that $\|V-\tilde{V}\|<\delta$; then
$\|\Delta_V-\Delta_{\tilde{V}}\|< 2\delta$.
\end{lemma}

\proof
Suppose that $\|V-\tilde{V}\|<\delta$, fix $\eps$ and drop the second
argument of $V(\phi,\eps)$ for convenience. Then $|V(\phi)-\tilde{V}(\phi)|<\delta$ 
for all $\phi$. The triangle inequality implies that
\begin{equation}
\label{eq:dV2delta}
\left| (V(\phi'+\phi)-V(\phi'))
- (\tilde{V}(\phi'+\phi)-\tilde{V}(\phi'))\right|< 2\delta
\end{equation}
for all $\phi'$ and $\phi\leq 1-\phi'$. For a given value of $\phi$ there is a
sequence $(s_n)_{n\in \mathbb{N}}$ such that $V(s_n+\phi)-V(s_n)\rightarrow
\Delta_V(\phi)$
as $n \rightarrow \infty$. By considering $\phi'=s_n$ in (\ref{eq:dV2delta}) we
can conclude that $\Delta_{\tilde{V}}(\phi)< \Delta_{V}(\phi)+2\delta$.
Similarly by considering $(\tilde{s}_n)_{n\in \mathbb{N}}$ such
that $\tilde{V}(\tilde{s}_n+\phi)-\tilde{V}(\tilde{s}_n)\rightarrow
\Delta_{\tilde{V}}(\phi)$ as $n \rightarrow \infty$ we have using
(\ref{eq:dV2delta}) that
$\Delta_{V}(\phi)< \Delta_{\tilde{V}}(\phi)+2\delta$. This
implies $\left| \Delta_V(\phi) - \Delta_{\tilde{V}}(\phi) \right| < 2 \delta$
for all
$\phi$ and $\eps$.
\qed

\begin{figure}
\begin{center}
\includegraphics[width=6cm]{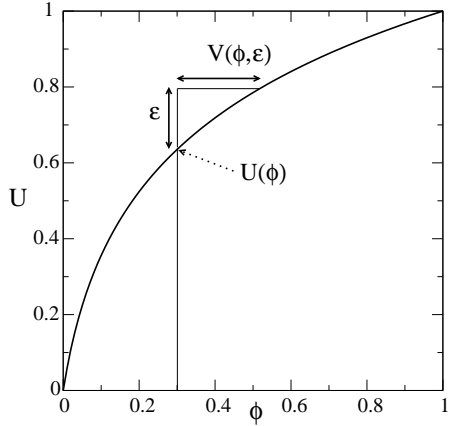}
\end{center}

\caption{
Plot showing $U(\phi)$ satisfying the conditions for Lemma~\ref{lemma:UV}
that hence giving a resetting function
$V(\phi,\eps)$ with increasing response. $U(\phi)$ in this case
is defined in ({\protect \ref{eq:Ub}}) with $b=3$.
\label{fig:VandUplots}
}
\end{figure}

The resetting rule
\begin{equation}
\phi\mapsto U^{-1}(U(\phi)+\eps)\label{eq:phi_Uphi}
\end{equation}
with a twice continuously differentiable 'potential' function $U$ was considered
 previously \cite{mirollo,ernst95,ernst98,timmeprl,timmechaos,timmephd} to
describe 
a phase jump induced by a reception ($R_{ij}$) of a pulse of strength $\eps$.
This `Mirollo-Strogatz' case is a special case of the more general
reception rule we use here, if we substitute
\begin{equation}
V(\phi,\eps)=\VMS(\phi,\eps) = U^{-1}(U(\phi)+\eps)-\phi. \label{eq:VfromU}
\end{equation}
We now show that, if $U$ is increasing concave downwards, this fits into the
more general reception scheme of Definition~\ref{def:cV}.
In Figure~\ref{fig:VandUplots} we illustrate a typical choice of $U$ and its
consequence for $V$ for given $\phi$ and $\eps$.

\begin{lemma} \label{lemma:UV}
Consider a response defined by $\VMS$ in (\ref{eq:phi_Uphi},\ref{eq:VfromU})
where  $0<\eps<c$ and $U$ is an unbounded twice continuously differentiable
function $U:[0,\infty)\rightarrow[0,\infty)$ that satisfies $U(0)=0$, $U(1)=1$, 
$0<c_1<U'(\phi)<C_1$  and $-C_2<U''(\phi)<-c_2<0$ for $\phi\in[0,1]$ 
where $c_k,C_k$ ($k=1,2$) are positive constants. Then $\VMS \in \cV$ for small
enough $c$.
\end{lemma}

Note that the details of the definition of $U$ are only critical
in the range $[0,1]$. This is because any time that $U>1$ we will get
resetting; the definition on $(1,\infty)$ is simply for 
expositional convenience.

\proof
One can directly verify that the conditions (i)--(ii) in the definition
of $\cV$ are satisfied. To show (iii), note that $U$ monotonic increasing and
surjective
on its range means it has a unique inverse $U^{-1}$ that is also monotonic
increasing.
To see that (iii) also holds, first observe that
\begin{equation}
\label{eq:Vprime}
\partial_{\phi}V(\phi,\eps) =
\frac{U'(\phi)}{U'( U^{-1}(U(\phi)+\eps))} -1.
\end{equation}
The condition $U'>0$ means that $U$ is strictly monotonic increasing.
The condition $U'<C_1$ means that $U(\phi+x)< U(\phi)+C_1 x$ for all $x$
satisfying $0< x \leq 1-\phi$. Together with the monotonicity of $U^{-1}$
we obtain, substituting $x=\eps/C_1$,
\begin{equation}
\label{eq:Ujumpbounded}
U^{-1}(U(\phi)+\eps) > \phi+ \frac{\eps}{C_1}.
\end{equation}
Similarly, $U''<0$ means that $U'$ is monotonic decreasing; $U''<-c_2$ implies 
$U'(\phi) > c_2 \eps/C_1 + U'(\phi+\eps/C_1)$ for all $\phi$ and thus, using
(\ref{eq:Ujumpbounded}),
\begin{equation}
U'(\phi) > U'(U^{-1}(U(\phi)+\eps)) + \frac{c_2}{C_1}\eps.
\end{equation}
Substituting into (\ref{eq:Vprime}) this leads to
\begin{equation}
\partial_{\phi}V(\phi,\eps) =\frac{U'(\phi)- U'(U^{-1}(U(\phi)+\eps))}{
U'(U^{-1}(U(\phi)+\eps))} > \frac{c_2}{C_1^2}\eps>0.
\end{equation}
Hence for fixed $\eps>0$ we have
\begin{equation}
\left|V(\phi'+\phi,\eps)-V(\phi',\eps)\right| = \left |\int_{\phi'}^{\phi'+\phi}
(\partial_{\phi} V)(s,\eps) \,ds \right| > \eps\frac{c_2}{C_1^2}\phi >0
\end{equation}
for all $\phi>0$, and so (iii) holds.
Finally, to show (iv), note that for $V=V_{\textsf{MS}}$ Eq.~(\ref{eq:phim}),
which
determines $\phim$, becomes $U(\phim)+\eps=1$ so that by definition
(\ref{eq:phim}), the
maximum change at a reset is
\begin{equation}
\label{eq:VoneMS}
\Vmax=1-U^{-1}(1-\eps).
\end{equation}
Clearly, $V_{\textsf{MS}}(\phi,\eps)>0$ for all $\eps>0$. Moreover, $\Vmax <1$
by definition (Eq.~(\ref{eq:VoneMS})),
\begin{equation}
\partial_{\eps}\Vmax=
\frac{1}{U'(U^{-1}(1-\eps))}>0
\end{equation}
and
\begin{equation}
\partial_{\eps \eps}\Vmax=
\frac{U''(U^{-1}(1-\eps))}{\left[U'(U^{-1}(1-\eps))\right]^3}<0
\end{equation}

because $U'>0$ and $U''<0$ in the relevant range. Thus $\Vmax$ is increasing and
concave and so we have the linear upper bound
$\Vmax \leq V_1 \eps$ where $V_1=\partial_{\eps} V_{\max}(0)$.
\qed

\subsection{Finite dimensional dynamics in finite time}

We proceed to show that the system of pulse coupled oscillators with increasing response
will reduce to a finite dimensional system for an open set of parameter values. 
Theorem~\ref{thmfinite} gives an estimate $k$ for the largest number of `past
firings' one needs to consider in the limit of large time. 
We can effectively reduce to a system of dimension at most $N(k+1)$ after a finite time.
The proof works by considering the maximum number of resets that can occur to an 
oscillator in any time interval $[0,\tau]$. This depends on the number
of firings that occurred in the previous time interval of the same length, giving 
the recurrence inequality in Lemma~\ref{lemestimate}.

We reduce the dynamics from a hybrid delay differential equation with infinite dimensional phase space
to a finite dimensional system by restricting to phase histories of the form:
\begin{equation}
\cP_{k}^{N}=\{\vtphi\in\cP^{N}~:~\s_{j,k+1}>\tau
\textrm{ for all }
j\in\{1,\ldots,N\}
\textrm{ for which a }
\sigma_{i,k+1} \textrm{ exists}
\}
\label{eq:kstatespace}
\end{equation}
namely, those that have at most $k$ firings within the delay time $\tau$.

\begin{lemma}\label{lem:finite}
Suppose that $\vtphi$ and $\vtpsi$ are in $\cP_{k}^N$ and $\Pi_k\vtphi=\Pi_k\vtpsi$.
Then
\begin{equation}
\Pi_k\Phi_t(\vtphi)=\Pi_k\Phi_t(\vtpsi)
\end{equation}
for all $t\geq 0$.
\end{lemma}

\proof
This result states that any two initial states
in $\cP_{k}^N$ whose past $k$ firings and phases are the same will have precisely
the same phases (and hence firings)  at all times in the future. This is because by definition all 
firings beyond the $k$th in the past will not affect the future phases. 
\qed

The previous Lemma does not necessarily imply that $\Phi_t(\vtphi)\in\cP_{k}^N$.

\begin{theorem}
\label{thmfinite}
Suppose that \( V \in \cV\). Then there is a \(k\geq 1\) (depending on $\eps$, 
$\tau$ and $V$) such that for any \(\vtphi \in \cP^{N}\)
there is a finite \(T>0\) (depending on \(\vtphi\), \(\eps\) and \(V\)) such
that $\Phi_{t}(\vtphi)\in \cP^N_k$ for all $t>T$. Moreover, after time $T$ the 
number of firings of an oscillator that can appear in any time interval of
length \(\tau\) is at most
\begin{equation}\label{eq:estk}
k = \left\lfloor \frac{1+\tau}{1-\Vone \eps} \right\rfloor
\end{equation}
where $\lfloor x\rfloor$ is the smallest integer less than or equal to $x$.
\end{theorem}

For any $k$ such that the conclusion of Theorem~\ref{thmfinite} holds we
say the dynamics {\em reduces} to $\cP_k^N$. We prove Theorem~\ref{thmfinite}
using the following Lemma. For any initial condition $\vtphi\in\cP^{N}$ let
$ \vtphi^{(m)}=\Phi_{m\tau}(\vtphi)$
($m \in\{ 0,1,2,\cdots\}$) and suppose that the number of firings
of $\vtphi^{(m)}$ in the interval $[-\tau,0)$ is $k_{m}$.

\begin{lemma}\label{lemestimate}
For all \(m\in\N\) we have
$k_{m} \leq \lfloor 1 + \tau + k_{m-1} \Vone \eps \rfloor$
\end{lemma}

\proof
Suppose that there have been at most $k_0$ firing events of each of the $N$
oscillators
during the time interval $[-\tau,0)$. Consider the events that affect the phase
$\phi_i$ 
of an arbitrary oscillator $i\in \{1,\ldots,N\}$ in the future time interval
$[0,\tau)$; 
there
will be $n_0(i)$ events in $[0,\tau)$ where at each event $N_n$, $N-1\geq
N_n\geq 1$ pulses 
of the oscillators
$\phi_j$, $j\neq i$ are simultaneously received by $i$. The total number of
received pulses is
$\sum_{n=1}^{n_0(i)} N_n \leq k_0 (N-1)$. The total change $\Delta\phi_i$ 
in phase of oscillator $i$ up to time $\tau$ is hence at most
\begin{equation}
\Delta\phi_i\leq\tau+V_1 \sum_{n=1}^{n_0} \frac{N_n}{N-1}\epsilon
\leq \tau + k_0 \Vone \eps.
\end{equation}
Hence  oscillator $i$ will have fired at most
\begin{equation}
k_{1}=\lfloor 1+\tau+k_{0}\Vone\eps \rfloor
\end{equation}
times in the interval $[0,\tau)$. Since $i$ is arbitrary, we have in general at 
most $k_m\leq \lfloor 1+\tau+k_{m-1}\Vone \eps \rfloor $ firing events of each 
oscillator in the interval $[\tau,0]$ and the Lemma follows.
\qed

\proof {\textbf{(of Theorem~\ref{thmfinite})}} Since $ \Vone \eps < 1$ applying 
Lemma~\ref{lemestimate} inductively we have \begin{eqnarray}
k_{m} & \leq & k_{0}(\Vone \eps )^{m}+(1+\tau)((
\Vone\eps)^{m-1}+(\Vone\eps)^{m-2}
+\cdots+(\Vone\eps)+1)\\
 & \leq & k_{0}(\Vone\eps)^{m}+\frac{1+\tau}{1- \Vone\eps}
\label{eq:timeestimate}
\end{eqnarray}
and so because $( \Vone\eps)^{m}\rightarrow0$, after some finite
time $T=m\tau$ (which can be estimated from (\ref{eq:timeestimate}) 
given $k_{0}$) we have
\begin{equation}
k_{m}\leq k=\left\lfloor \frac{1+\tau}{1-\Vone \eps}\right\rfloor.
\end{equation}
The remainder of the theorem is a consequence of applying Lemma~\ref{lem:finite}.
\qed

The estimate of $k$ will typically be a very poor upper bound. If the number of
oscillators $N$ is small one can improve on these bounds by noting that the
pulses are `quantized' into multiples of $\eps/(N-1)$. In particular, one can
replace the upper bound (\ref{eq:Vone}) by requiring only that there is a $V_1$ such that
\begin{equation}
V\left(\phim,\frac{k\eps}{N-1}\right) \leq V_1 \frac{k\eps}{N-1} \leq 1
\end{equation}
for all $1\leq k \neq (N-1)$. For larger $N$ this approaches (\ref{eq:Vone}).

As a consequence of Lemma~\ref{lemestimate} we can now justify 
defining the dynamics as a semiflow on $\cP^{N}$.

\begin{corollary}
A system with $V\in\cV$ that satisfies $\Vone\eps<1$ will remain free of
Zeno states for any $\eps>0$ and $\tau>0$.
\end{corollary}

This is because, although the number of firings in an interval $[0,\tau]$
can increase if $\Vone \eps$ is large, the maximum rate is given in
Lemma~\ref{lemestimate} and there cannot be infinitely many firings in any time 
interval of finite length.
Another consequence of the proof of Theorem~\ref{thmfinite} is that the
semiflow on $\cP^N$ has a restriction to $\cP^N_k$ that is a well-defined
semiflow.

\subsection{$2N$ dimensional state space in finite time ($k=1$)}

In the case that
\begin{equation}
1+\tau<2-2\Vone\eps,\label{eq:k=1}
\end{equation}
we can apply Theorem~\ref{thmfinite} to see that $1\leq(1+\tau)/(1-\Vone\eps)<2$,
and so $k=1$. In fact, this is not a very tight estimate, and in
many cases where (\ref{eq:k=1}) does not hold the dynamics still
reduces to $k=1$.
For a particular initial condition $\vtphi$ one can practically check
the dimension it reduces to by finding the $\limsup$ of the number of firings
of one oscillator in any time interval of length $\tau$ for a neighbourhood of that
initial condition.

If at most one previous firing is needed to determine the phases
uniquely in the future, the dynamics reduces to a semiflow on
a $2N$ dimensional space
$$
\Pi_1\left(\cP_1^N\right)
=\{(\phi_{1},\cdots,\phi_{N},\s_{1},\cdots,\s_{N})~:~~\phi_{i}\in[0,1),
~\s_{i}\in[0,\infty)\}
$$
where we write $\s_i=\s_{i,1}$ in shorthand. For the oscillators we consider, the semiflow
\begin{equation}
\Phi_{t}:\cP_1^N \rightarrow \cP_1^N
\end{equation}
for $t\geq0$ is well defined for $\eps$ and $\tau$ small enough;
Appendix~\ref{App:k=1} gives an explicit form on this reduced space.

\section{Unstable attractors for pulse-coupled oscillators}
\label{secuasforpulsecos}

The main result in this section is a proof that unstable
attractors exist and are robust in the system of delay pulse
coupled oscillators discussed in Section~\ref{secpulsecos}.

\subsection{Unstable Attractors: Existence and Robustness}

For this section we consider the potential function
\begin{equation}
U(\phi)=U_b(\phi)~\mbox{ where }~U_b(\phi)=\frac{1}{b}\ln(1+(e^{b}-1)\phi)
\label{eq:Ub}
\end{equation}
used in \cite{timmeprl}. Computing the associated response function $V$ defined
by 
(\ref{eq:VfromU}) we find an affine relationship between $\phi$ and $V$:
\begin{equation}
V(\phi,\epshat)=\frac{e^{\epshat b}-1}{e^b-1} + \left(e^{\epshat b}-1\right)
\phi
\end{equation}
and one can check that $V\in\cV$; this is a particular case of
Lemma~\ref{lemma:UV}; the relationship between $U$ and $V$ is
illustrated in Figure~\ref{fig:VandUplots}.

\begin{theorem}\label{thm:uasexist}
Consider the system of oscillators defined in Section \ref{sect:defsemiflow}
with \(U\) 
defined by (\ref{eq:Ub})
with \(N=4\), \(\tau=0.14\) and \(\eps=0.24\) and $b=3$. This system has a
periodic orbit 
that is an unstable attractor with zero measure local basin.
\end{theorem}

\proof
By applying Theorem~\ref{thmfinite} one can directly verify that any initial 
condition $\vtphi\in\cP^4$ will reduce to the $4(1+4)=20$ dimensional space $\Pi_4(\cP_4^4)$ 
after a finite time, i.e. such that only the phase and last four firings of each oscillator 
contribute to the dynamics at any point in the future. Now consider the open 
set of initial conditions defined by
\begin{equation}
(\phi_{1},\cdots,\phi_{4},\s_1,\cdots,\s_4)\in C_0
\end{equation}
where
\begin{equation}
\label{eq:C0}
C_{0}=(0.7,0.7,0.3,0.95,0.5,0.5,0.5,0.5)+[0,0.001]^{8}\subset \Pi_{1} \left(\cP^4\right).
\end{equation}
As a slight abuse of notation, we will also write $C_0$ as meaning $\Pi_{1}^{-1}(C_0)$ if the
context is clear. Note that all the $\s_i>\tau=0.14$, meaning that $C_0\subset \cP_1^4$.
Theorem~\ref{thmfinite} means that at points in the future we may in principle arrive at
states that are in $\cP_k^4$ for $k=4$ but not for $k\leq 3$. However, we will verify below that 
$\Phi_{t}(C_0)\subset \cP_1^4$ for all $t>0$.

Note that $\Pi_{1}^{-1} C_0$ is an open set, as is its image under $\Pi_k$ is open for any
$k\geq 0$; we will show that it is in the basin of attraction of a saddle periodic orbit.  In what follows 
we write $C_{k}$ to denote the $k$th intersection of 
the trajectories in $C_{0}$ with one or more of the `events' $S_i$ or $R_{ij}$. Since
the information about the original initial condition of the $\s_i$ has been lost we 
list only the first four components. We compute that
\begin{equation}
C_{1}=[0.749,0.751]\times[0.749,0.751]\times[0.349,0.351]\times\{0\}
\label{eq:C1}
\end{equation}
due to the first event being $S_{4}$. The next event is $R_{4}$ which will give
\begin{equation}
C_{2}\subset\{(0,0,\phi_3,0.14)~:~~\phi_3\in[0.63,0.64]~~\}\label{eq:C2}
\end{equation}
and so we have reduced to a
one dimensional subspace of phase space at this point. A very
tedious direct numerical verification listed in Table~\ref{tab:C0traj}
shows that {\em all} of the points starting in $C_{2}$ after
a further 32 events reach a single point
\begin{equation}
\begin{array}{l}
C_{34}=\{P\}~~\mbox{ with }~~P:=(0,0,a,a,0,0,b,b)
\end{array}
\label{eq:C22P}
\end{equation}
where $a\approx 0.62307$ and $b\approx 0.57088$ can be computed exactly
(see Appendix~B). 
We present in Figure~\ref{fig_return} the values
of the coordinates depending on choice of $\phi_3$ over a larger
range than in $C_{2}$; observe that a one dimensional set of possible initial
values converges to exactly the same point.
The point $P$ (\ref{eq:C22P}) is 
a cross section to a periodic orbit where two pairs
of oscillators remain synchronized. 
Since the periodic orbit through $P$ contains the open set $C_0$ in its basin
it is a Milnor attractor for any reduction of the system to finite dimensions, 
for example to $\cP_4^4$.  

\begin{table}
{\small
$$
\begin{array}{r|cccccccc}
k & \phi_1 & \phi_2 & \phi_3 & \phi_4 & \sigma_1 & \sigma_2 & \sigma_3 &
\sigma_4 \\
\hline
0 & 0.70000 & 0.70000 & 0.30000 & 0.95000 & 1.00000 & 1.00000 & 1.00000 &
1.00000 \\
1 & 0.75000 & 0.75000 & 0.35000 & 0.00000 & 1.05000 & 1.05000 & 1.05000 &
0.00000 \\
2 & 0.00000 & 0.00000 & 0.63712 & 0.14000 & 0.00000 & 0.00000 & 1.19000 &
0.14000 \\
3 & 0.19219 & 0.19219 & 0.00000 & 0.48478 & 0.14000 & 0.14000 & 0.00000 &
0.28000 \\
4 & 0.43650 & 0.43650 & 0.14000 & 0.80846 & 0.28000 & 0.28000 & 0.14000 &
0.42000 \\
5 & 0.62804 & 0.62804 & 0.33154 & 0.00000 & 0.47154 & 0.47154 & 0.33154 &
0.00000 \\
6 & 0.99058 & 0.99058 & 0.61365 & 0.14000 & 0.61154 & 0.61154 & 0.47154 &
0.14000 \\
7 & 0.00000 & 0.00000 & 0.62307 & 0.14942 & 0.00000 & 0.00000 & 0.48095 &
0.14942 \\
8 & 0.19219 & 0.19219 & 0.00000 & 0.50000 & 0.14000 & 0.14000 & 0.00000 &
0.28942 \\
9 & 0.43650 & 0.43650 & 0.14000 & 0.82781 & 0.28000 & 0.28000 & 0.14000 &
0.42942 \\
10 & 0.60870 & 0.60870 & 0.31219 & 0.00000 & 0.45219 & 0.45219 & 0.31219 &
0.00000 \\
11 & 0.96599 & 0.96599 & 0.58906 & 0.14000 & 0.59219 & 0.59219 & 0.45219 &
0.14000 \\
12 & 0.00000 & 0.00000 & 0.62307 & 0.17401 & 0.00000 & 0.00000 & 0.48620 &
0.17401 \\
13 & 0.19219 & 0.19219 & 0.00000 & 0.53974 & 0.14000 & 0.14000 & 0.00000 &
0.31401 \\
14 & 0.43650 & 0.43650 & 0.14000 & 0.87833 & 0.28000 & 0.28000 & 0.14000 &
0.45401 \\
15 & 0.55817 & 0.55817 & 0.26167 & 0.00000 & 0.40167 & 0.40167 & 0.26167 &
0.00000 \\
16 & 0.90176 & 0.90176 & 0.52483 & 0.14000 & 0.54167 & 0.54167 & 0.40167 &
0.14000 \\
17 & 0.00000 & 0.00000 & 0.62307 & 0.23824 & 0.00000 & 0.00000 & 0.49990 &
0.23824 \\
18 & 0.19219 & 0.19219 & 0.00000 & 0.64354 & 0.14000 & 0.14000 & 0.00000 &
0.37824 \\
19 & 0.43650 & 0.43650 & 0.14000 & 0.00000 & 0.28000 & 0.28000 & 0.14000 &
0.00000 \\
20 & 0.74709 & 0.74709 & 0.37016 & 0.14000 & 0.42000 & 0.42000 & 0.28000 &
0.14000 \\
21 & 0.00000 & 0.00000 & 0.62307 & 0.39291 & 0.00000 & 0.00000 & 0.53291 &
0.39291 \\
22 & 0.19219 & 0.19219 & 0.00000 & 0.89350 & 0.14000 & 0.14000 & 0.00000 &
0.53291 \\
23 & 0.29869 & 0.29869 & 0.10650 & 0.00000 & 0.24650 & 0.24650 & 0.10650 &
0.00000 \\
24 & 0.43650 & 0.43650 & 0.14000 & 0.05679 & 0.28000 & 0.28000 & 0.14000 &
0.03350 \\
25 & 0.70451 & 0.70451 & 0.32758 & 0.16330 & 0.38650 & 0.38650 & 0.24650 &
0.14000 \\
26 & 0.00000 & 0.00000 & 0.62307 & 0.45879 & 0.00000 & 0.00000 & 0.54199 &
0.43549 \\
27 & 0.19219 & 0.19219 & 0.00000 & 0.99996 & 0.14000 & 0.14000 & 0.00000 &
0.57549 \\
28 & 0.19222 & 0.19222 & 0.00004 & 0.00000 & 0.14004 & 0.14004 & 0.00004 &
0.00000 \\
29 & 0.43650 & 0.43650 & 0.14000 & 0.19214 & 0.28000 & 0.28000 & 0.14000 &
0.13996 \\
30 & 0.56917 & 0.56917 & 0.19224 & 0.19218 & 0.28004 & 0.28004 & 0.14004 &
0.14000 \\
31 & 0.00000 & 0.00000 & 0.62307 & 0.62301 & 0.00000 & 0.00000 & 0.57087 &
0.57083 \\
32 & 0.19219 & 0.19219 & 0.00000 & 0.00000 & 0.14000 & 0.14000 & 0.00000 &
0.00000 \\
33 & 0.56912 & 0.56912 & 0.19219 & 0.19219 & 0.28000 & 0.28000 & 0.14000 &
0.14000 \\
34 & 0.00000 & 0.00000 & 0.62307 & 0.62307 & 0.00000 & 0.00000 & 0.57088 &
0.57088 \\
\end{array}
$$
}
\caption{
\label{tab:C0traj}
The state of the system at times of the $k$th event on a trajectory
starting with an initial condition
in $C_0$ (see equation (\ref{eq:C0}) in the proof of
{\protect Theorem~\ref{thm:uasexist}}) the phases $\phi_i$ and firing
times $\sigma_i$ defining the state at the event times are shown
to five decimal places. All initial conditions in $C_0$
finally arrive at the same point in $C_{34}$ because of the firing-induced
resetting at stages $k=2$ and  $k=32$.
}
\end{table}

\begin{figure}
\begin{center}
\includegraphics[width=9cm]{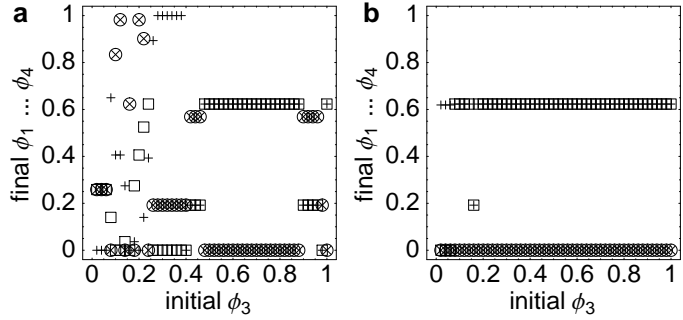}
\end{center}

\caption{
\label{fig_return}
Dependence of dynamics on the initial phases $(0,0,\phi_3,0.14) \in C_2$. Panel (a)
shows the phases ($\phi_{1}$($\bigcirc$),
$\phi_{2}$($\times$), $\phi_{3}$ ($\square$), $\phi_{4}$($+$))  after $32$
further intersections with discontinuities of the semiflow (`events')
in dependence of $\phi_3$.
Observe that there is a large range of $\phi_3$ for which the dynamics arrives
at precisely the same
point. Panel (b) shows the actual attractor reached from the above initial
conditions (shown after the $52^{nd}$ reset of oscillator $i=1$. Surprisingly,
the same attractor is reached by several initial conditions whose trajectory
after 34 events differs substantially.
}
\end{figure}

Now we show that the orbit through $P$ is unstable. 
In order to do this we show that almost all
perturbations
of the first two coordinates grow to finite size under the action of the
local return map. A complete stability analysis is presented in the
Appendix \ref{App:unstable}. In particular, as is shown below, this proves
instability in general,
not only linear instability.

In this analysis we consider the return map from the plane $\phi_{2}\equiv0$
to itself. The pulses $\s_i$ are kept locked to the phases $\phi_i$ such
that 
the nonlinear return map (\ref{eq:F(delta)})
\begin{equation}
\boldsymbol{\delta}'=\mathbf{{F}}(\boldsymbol{\delta})\label{eq:Fdelta}
\end{equation}
is three-dimensional and maps a perturbation
$\boldsymbol{\delta}:=(\delta_{1},\delta_{3},\delta_{4}):=
\vphi(0)-\vphi^*(0)$
to the periodic orbit started at (\ref{eq:C22P}).

\begin{equation}
\boldsymbol{\phi}^{*}(0)=(0,0,a,a)
\end{equation}
to the perturbation $\boldsymbol{\delta}'$ when the plane
$\phi_{2}=0$ is reached the next time.

If we take a general perturbation (satisfying $\delta_1\neq \delta_2$) just
before arriving at $(0,0,a,a)$,
observe that the $\sigma_i$ are all larger than
$\tau$ and hence will not affect the future of the
phases. Starting with such a perturbation,
when it hits the next event we can arrive at an arbitrary
perturbation satisfying $\delta_{1}>\delta_{2}=0$ such that
immediately after the perturbation, events occur in the
order $S_{1},S_{2},(S'_{3},S'_{4})$.
Note that the order of the events $S_{3}$ and $S_{4}$ does not play a role as
they become
synchronized at their first firing.

As demonstrated explicitly in Appendix B, there is a small
$\Delta^*>0$ such that the return map (\ref{eq:Fdelta}) defined in
the three-dimensional region
\begin{equation}
\cR=\{\boldsymbol{\delta}~:~\delta_1>\delta_2=0 \mbox{ and } |\delta_i|<\Delta^*
\mbox{ for all } i \in \{1, 3, 4\} \}
\end{equation}
is given by a smooth map with $\mathbf{F}(\mathbf{0})=\mathbf{0}$ (and
such that $\mathbf{F}(\cR)$ is a line that intersects $\cR$).
In particular, given the map $\mathbf{F}(\boldsymbol{\delta})$
in terms of an arbitrary function $V\in\cV$
(see Appendix B) we analyse of the fully nonlinear local dynamics of a
sufficiently small perturbation $\boldsymbol{\delta}=
(\delta_1,0,0)$, $0 < \delta_1 < \Delta^*$, to the  periodic orbit. For the
following estimate, we write $\Delta_\eps(\phi)$ for $\Delta_V(\phi,\eps)$ in
shorthand notation.
First note that requirement (iii) in Def. \ref{def:cV} implies that there is a
$\Delta_\eps(\delta)$ such that for all $\delta>0$, 
$H_{\eps}(\phi+\delta)-H_{\eps}(\phi)\geq \delta + \Delta_\eps(\delta)$ for all
$\eps>0$, in particular, $H_{\eps}$ is monotonic increasing in $\phi$. This
implies that

\begin{equation} \left.
\begin{array}{lllll}
F_{1}(\boldsymbol{\delta}) & = & H_{2\epshat}(H_{\epshat}(\tau+\delta_{1})+\tau)
& - & H_{2\epshat}(H_{\epshat}(\tau-\delta_{1})+\tau+\delta_{1})\\
& \geq &
 H_{2\epshat}(H_{\epshat}(\tau)+\delta_{1}+\Delta_{\epshat}(\delta_1)+\tau)
 & - & H_{2\epshat}(H_{\epshat}(\tau)-\Delta_{\epshat} (\delta_{1})+\tau)\\
& \geq &
 \delta_1 + \Delta_{\epshat} (\delta_1)+
\Delta_{2\epshat}(\delta_1+\Delta_{\epshat} (\delta_1))
 & + & \Delta_{\epshat} (\delta_1) + \Delta_{2\epshat}(\Delta_{\epshat}
(\delta_1)) \\
& > & \delta_1 + 2 \Delta_{\epshat} (\delta_1).
\end{array}\right.
\label{eq:A:F_i_bis}
\end{equation}

Hence we find that $F_1(\boldsymbol{\delta})>\delta_1+ 2 \Delta_{\epshat}
(\delta_1)$ implies that any perturbation $\boldsymbol{\delta}=(\delta_1,0,0)$
satisfying $\delta_1>0$ increases (e.g. in maximum norm). Because the function
$\Delta_{\epshat} (\delta)$ is non-decreasing in $\delta$, we see that there is
a $\Delta^*>0$ for which any sufficiently small perturbations (satisfying
$|\delta_i|<\Delta^*$) leave the region $\cR$ in finite time $t_\cR$, bounded by
\begin{equation}
t_\cR \leq \frac{\Delta^*-\delta_1}{2\Delta_{\epshat} (\delta_1)}\ .
\end{equation}

By the permutation symmetry of the system, the analysis in the case
$\delta_1<\delta_2=0$ is analogous.
This means that almost all sufficiently small perturbations
$\boldsymbol{\delta}$ leave a neighbourhood of the attractor in finite time,
proving that the attracting periodic orbit considered is also unstable.
The case $\delta_2<\delta_1$ can
be treated analogously by the symmetry of the globally coupled network.
Taken together there is a neighbourhood
$\cR'$ of $\boldsymbol{\delta}=\mathbf{0}$ such that
all points starting in
$\cR'$ with $\delta_1\neq 0=\delta_2$ will iterate away until they leave this
neighbourhood. This proves instability of the orbit through $P$ (\ref{eq:C22P}).
\qed

\vspace{3mm}

\begin{figure}
\begin{center}\includegraphics[width=7cm]{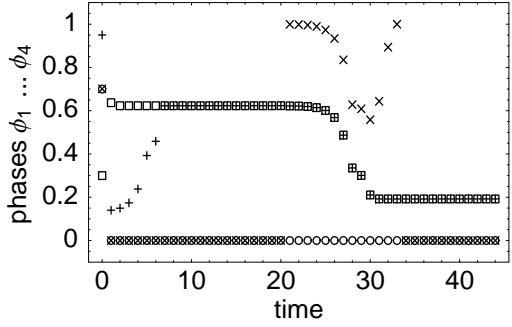}
\end{center}
\caption{
A four oscillator network ($N=4$, $\eps=0.24$, $\tau=0.14$, $U=U_{b}$, $b=3$)
exhibiting an unstable attractor. At time zero, initial phases
in $C_0$ are shown. The time axis is discrete and shows the phase when
oscillator $i=2$ fires. Just after
the firing event $S_{2}$ of oscillator $i=2$ at times $t_{n}$,
all phases $\phi_{i}(t_{n})$ are plotted ($\phi_{1}$($\bigcirc$),
$\phi_{2}$($\times$), $\phi_{3}$ ($\square$), $\phi_{4}$($+$)).
There is convergence of the trajectory in state space in finite time
towards the periodic orbit passing through ({\protect \ref{eq:C22P}}).
An instantaneous small random perturbation ($|\delta_{i}|<10^{-3}$) to the
phases at time $n=20$ leads to a switching to another attractor, 
demonstrating instability of the original periodic orbit.
\label{fig:XC}}
\end{figure}

Figure~\ref{fig:XC} shows the dynamics starting from one initial condition
(\ref{eq:C0})
used in the proof of the theorem. The convergence in finite time to
the attractor is seen. A small random perturbation to the phases of all
oscillators is
expanded by the instability, yielding convergence towards another attractor.

\begin{theorem}\label{thm:uasrobust}
The unstable attractors of Theorem~\ref{thm:uasexist} are robust in
the sense that there is an open neighbourhood in $\cV$ of
$V(x,\epshat)=H_{\epshat}(x)-x$ given by (\ref{eq:VfromU},\ref{eq:Ub})
and an open neighbourhood of $(\tau,\eps)$ such that a system defined by any
$\tilde{V}$, $\tilde{\tau}$, and $\tilde{\eps}$ in these neighbourhoods, has an
unstable attracting periodic orbit that remains
close to that of Theorem~\ref{thm:uasexist}.
\end{theorem}

\proof
We refer to the return map $\mathbf{F}$ (\ref{eq:F(delta)}) derived in the
Appendix~B
and already discussed in the proof of Theorem~\ref{thm:uasexist}.
Note that in the derivation of the return map $\mathbf{F}$, the form
of the transfer function $H$  and the precise values of the parameters $\tau$
and $\eps$ are not important, because $\mathbf{F}$ continuously depends on $H$,
$\tau$, and $\eps$ and changes its form only if the sequence of events
occurring along the trajectory on or near the periodic orbit is altered.
Thus, an equivalent periodic orbit exists if the dynamical system is slightly
perturbed, i.e. if $H$, $\tau$, and $\eps$, are slightly changed from their
original values. In particular, one can choose
$H_{\eps}(\phi)=\phi+V(\phi,\eps)$, where $V$ is sufficiently close to the
original. Since the mechanism of dimensional reduction and  attractivity depends
only on the supra-threshold
synchronization mechanism given by the induced firing events $S'_i$,
the slightly perturbed periodic orbit will also be an attractor.
Moreover, for a sufficiently small perturbation to $V$, $\tau$, and $\eps$ we 
still obtain
\begin{equation}
F_1(\boldsymbol{\delta})\geq 2 \Delta_{\epshat}(\delta_1)+\delta_1 
\mbox{ if } \Delta^*>\delta_1>0 ,
\end{equation}
which implies instability.
(cf. the proof of the previous theorem).
Again, since the return map $\mathbf{F}$ does not change form if the parameters 
$H$, $\tau$, and $\eps$ are slightly perturbed, also the proof of instability is
independent of these parameters of the system. This completes the proof.
\qed

This result implies that qualitatively the same dynamics is obtained by any
flow defined by a $\tilde{V}$ nearby the original $V$ (\ref{eq:VfromU});
in particular it does not have to be defined in terms of a  
any potential function $U$ by (\ref{eq:VfromU}).

\subsection{Interpretation in terms of dimension jump}

\begin{figure}
\begin{center}\includegraphics[width=10cm]{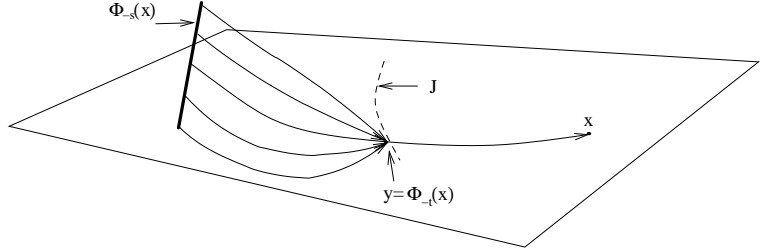}
\end{center}
\caption{
Schematic showing a point $x$ that has a unique inverse $\Phi_{-u}(x)$ for
$0<u\leq t$. If we define $y=\Phi_{-t}(x)$ then $y$ is in $J$ the jump
set (in this case $j(y)=1$) and for any $t<s$ the dimension
of the set $\Phi_{-s}$ is greater than zero.
\label{fig_jumpset}}
\end{figure}

Since the proof of Theorem~\ref{thm:uasexist} is rather involved,
we discuss some intuition in terms of what we term a dimension jump 
of the semiflow.  Given a semiflow $\Phi_{t}:M\rightarrow M$ that 
is not invertible, one can nevertheless define, for all $x \in M$, a set-valued
inverse 
for any $t>0$ by
\begin{equation}
\Phi_{-t}(x)=\{ y\in M~:~\Phi_{t}(y)=x\}
\end{equation}
and the semiflow can be extended uniquely to an invertible flow precisely
when $\Phi_{-t}(x)$ is a single point for all $x$ and $t>0$.
Given sufficient regularity on $\Phi$, for example sufficient to ensure that
 $\Phi_{-t}(x)$ is a union of manifolds, we can
compute its dimension. If all $\Phi_{-t}(x)$ are non-empty
then $\dim(\Phi_{-t}(x))$ will be a non-decreasing function of $t$.
The \emph{dimension jump}
\begin{equation}
j(x)=\lim_{t\rightarrow 0-}\left(\dim(\Phi_{t}(x))\right)
\end{equation}
at $x$ is of particular interest and we refer to
\begin{equation}
J=\{ x\in M~:~j(x)>0\}
\end{equation}
as the \emph{dimension jump set}, shown schematically
in Figure~\ref{fig_jumpset}. Such behaviour
can be found in systems exhibiting sliding modes
\cite{slidingref}, and we suggest that such systems may display
unstable attractors.

Observe that trajectories converging to an invariant set that
pass through $J$ may cause locally unstable invariant sets to be 
unstable attractors. For the system of delay coupled oscillators with $k=1$
there 
are two events on which dimension jump can occur:
\begin{enumerate}
\item
For one of the firing events $S_{i}$ or $S'_i$ we lose dependence on the delayed
phase $\s_{i}$. 
\item
At a resetting $R_{j}$ that induces firing in two oscillators $S'_{i_{1}}$ and
$S'_{i_{2}}$ we
synchronize the oscillators $i_{1}$ and $i_{2}$ beyond that point; this also
causes a dimension jump.
\end{enumerate}
In out system, these events have `jump sets' that are codimension one and
typically they cause a loss of one dimension; however when the oscillators
become synchronized, several such events can occur at once resulting in a loss
of more dimensions.

The proof of Theorem~\ref{thm:uasexist} works by choosing an open set of initial
conditions that reduces dimension down to one dimension after two further
discrete events. The final loss of dimension that reduces the open
set down to one point in the section occurs only a while later, and
this is because there is a global connection in phase space that we
need to follow.

\section{Discussion}
\label{secdiscuss}

In this paper we give a rigorous definition for a novel type of Milnor
attractor, 
the \emph{unstable attractor}, introduced in \cite{timmeprl,timmephd}. We
classify these attractors 
in terms of presence or absence of local basin. These attractors 
are surprising in many ways; notably they have `local repelling'
properties in that they have neighbourhoods such that almost all
points leave the neighbourhood after a finite time.

We show that unstable attractors not only exist, but even appear robustly in
certain
classes of hybrid dynamical systems of interest as physical and 
biological models; namely in networks of pulse-coupled oscillators.
The robustness that we have demonstrated is limited in the sense 
that we only consider perturbations that keep the time delay 
identical for all connections and the coupling
function identical for all oscillators. We do not consider 
perturbations that break the identical nature of the oscillators or 
the symmetry of the system. On introduction of such
general perturbations we expect no robustness for
such unstable attractors, although one may obtain attractors that
remain `nearby' to where a network of unstable attractors used to exist, i.e.
close to a collection of unstable periodic orbits and their heteroclinic
connections.

For perturbations that break only part of the symmetry,
unstable attractors may still persist. For instance there is numerical evidence
that 
unstable attractors can still exist even the network is not
coupled all-to-all, cf.~\cite{zumdieck}. 
The exact consequences of non-identical oscillators and more general
broken symmetry remains to be studied.

If we perturb the system in such a way that it becomes
smooth and invertible, the unstable attractors will
disappear; any saddle periodic orbit
will have a zero measure basin simply by existence of
a global stable manifold of dimension less than
that of phase space; however as indicated in
Section~\ref{secunstableatts} one should be able to
make the system arbitrarily differentiable and still obtain
unstable attractors, as long as non-invertibility is retained.

In Proposition~\ref{prop:nounsts} we have shown that certain 
conditions on a general dynamical system exclude the possibility of
unstable attractors with zero local basin. To generalize this finding 
to also exclude unstable attractors with positive local basin, we believe
it will be necessary to strengthen these conditions. Generalization of 
our result is not straightforward because a (local) basin may be highly
irregularly, as for riddled basin attractors.

Our proofs of the main results are not very
elegant; this is because they rely on global features of
the hybrid dynamical system, and we have not found them easily reducible
to simpler arguments. We believe however that it should be
possible for exact numerical methods \cite{timmechaos} to identify when unstable
attractors are present in such systems.

The numerical results in \cite{timmephd} suggest that networks
of unstable attractors exist in a wide range of systems of pulse-coupled 
oscillators. 
Preliminary studies have shown that the networks of unstable attractor may 
often be transitive for a wide range of parameters and system sizes. Revealing
the 
mechanisms underlying the existence, structure and prevalence 
of such networks of unstable attractors is a challenging task for future
research.

\textbf{Acknowledgements:}
PA thanks the Max-Planck-Institut f\"{u}r Dynamik und Selbstorganisation 
(formerly: MPI f\"{u}r Str\"omungsforschung)
for their hospitality and financial support, as well as a grant from
the Leverhulme Foundation for partial support. MT thanks
Leonid Bunimovich, Stefan Gro\ss kinsky and Raoul-Martin Memmesheimer for useful and interesting conversations relating to this work as well as the 
Department of Mathematical Science, University of Exeter for 
their hospitality and financial support.

\appendix

\section{Explicit form of the semiflow for $k=1$}
\label{App:k=1}
We give the semiflow of globally coupled oscillator networks
explicitly for the case $k=1$ by the following algorithm.
Consider
$(\phi,\sigma)=(\phi_{1},\ldots,\phi_{N},\s_{1},\ldots,\s_{N})\in \Pi_1(\cP_1^N)$ 
where the $i$th component represent the $i$th phases and the $n+i$th component
represents the time since the $i$th oscillator fired. Suppose that $\Phi_t(\cP_1^N)\subset \cP_1^N$
for all $t\geq 0$. We note that we can represent the flow on $\Pi_1(\cP_1^N)$ by
\begin{equation}
\Phi_{t}(\phi,\s)=(\phi,\s)+t(1,\cdots,1)
\end{equation}
for all $0<t$ such that $\phi_{i}<1$ and $\s_{i}<\tau$. After
time
\begin{equation}
\label{eq:algorT}
T(\phi,\s)=\mbox{smallest positive value in }
\left( 1-\phi_{i},\tau-\s_{i}~:~i\in \{ 1,\ldots,N \}\right)
\end{equation}
we hit the first event and apply a mapping according to the 
following algorithm:

\begin{itemize}
\item Set $(\phi,\s)\mapsto(\phi,\s)+T(\phi,\s)(1,\cdots,1)$.
\item For each $i$ such that $\s_{i}=\tau$ we say $R_{j}$ occurs for all $j\neq
i$ and set
$\phi_{j}(T)=\phi_{j}(T^-)+ V(\phi_{j},\epshat)$ for $\epshat$ defined as in
(\ref{eq:epshat}).
\item If there is an $i$ such that $\phi_{i}(T)\geq1$ then we say $S_{i}$ occurs
and set $\phi_{i}(T):=0$ and $\s_{i}(T):=0$.
\end{itemize}
After applying this algorithm we are assured that $T(\phi,\s)>0$, as
$\phi_{i}(T)\in[0,1)$ and $\s_{i}(T)\neq\tau$ for $i=1..n$. Thus we can take
this state as a new initial condition and rerun the algorithm sequentially from
event to event.

\section{Return map near unstable attracting periodic orbit
\label{App:unstable}}

For the main results of this paper (Theorems 2 and 3) we use a nonlinear return
map $\mathbf{F}(\boldsymbol{\delta})$
of perturbations $\boldsymbol{\delta}$ to the unstable attracting
periodic orbits in a network of $N=4$ oscillators using
the semiflow detailed in Appendix~\ref{App:k=1}. We
detail the derivation of this return map and analyze the periodic orbit 
that is the unstable attractor.

We choose $\phi_{2}(t_{0})=0$ where $t_{0}=0$, and consider cases where
the initial $\sigma_{ij}>\tau$.
First we analyse the dynamics on the periodic orbit. To establish
instability we then consider, without loss of generality, perturbations
to the periodic orbit phases at a point on the orbit where the time since 
last pulse is greater than $\tau$ for all oscillators. This means that
the future depends only on the current phases and we can effectively
consider $\phi_{i}(t)-\s_{i,k}(t)=\textrm{const}$ for all 
$i\in\{1,2,3,4\}$ and all
$t$ satisfying $t_1\leq \s_{i,k}(t)\leq t_2$ where $t_1$ defines the time 
of the last event such that $\s_{i,k}(t_1)=0$ and $t_2>t_1$ the time of 
the next event analogous to (\ref{eq:algorT}). Within this convention,
we consider the return map from the three-dimensional plane defined
by $\phi_{2}=0$ to itself and establish instability for this
return map to prove instability of the periodic orbit.

The following analysis applies for the parameters $\eps=0.24$, $\tau=0.14$,
and the function $U=U_{b}$ (\ref{eq:Ub}) where $b=3$.
It will turn out that it also applies for a neighbourhood thereof,
showing robustness of the instability of the periodic orbit.
For compact notation, we introduce the \emph{transfer function}
\begin{equation}
H_{\eps}(\phi)=V(\phi,\eps)+\phi .  \label{eq:A:H_epsV}
\end{equation}
This transfer function mediates the action of an incoming pulse of strength
$\eps$ on the phase $\phi$. 
We obtained the sequence of discrete events and the numerical values in the
following analysis for the specific choice (\ref{eq:VfromU},\ref{eq:Ub}).

\subsection{Periodic orbit dynamics}

First we determine the dynamics of the unperturbed orbit
(Table~\ref{Tab:unperturbed_PO}).
Throughout the table,
the phases after each event are labelled $p_{i,k}$ where the first
index labels the oscillator $i\in\{1,2,3,4\}$ and the second index
counts the events $k$ occurring for a \textit{perturbed} trajectory
(see Table \ref{Tab:perturbed_PO} below) starting with $k=0$ for
the initial condition (first row of Table \ref{Tab:unperturbed_PO}).
Since the second oscillators $i=2$ and $i=4$ in the two synchronous
groups have the same phases as the oscillators $i=1$ and $i=3$,
respectively, we define $p_{2,k}:=p_{1,k}\textrm{ and }p_{4,k}:=p_{3,k}$
for all $k$.
We show that the unperturbed orbit indeed is periodic with period
\begin{equation}
T=2\tau+1-p_{1,3}\end{equation}
 given by the time $t$ in the last row of Table \ref{Tab:unperturbed_PO}
where\begin{equation}
p_{1,3}=H_{2\epshat}(H_{\epshat}(\tau)+\tau)\end{equation}
 is also read off the table. Here we use the abbreviation $\epshat=\eps/(N-1)$
for the strength of an individual pulse.

\begin{table}
\begin{center}
\begin{tabular}{|l|l|cc|}
\hline
Event
&
$t$
&
$\phi_{1}(t)\equiv\phi_{2}(t)$
&
$\phi_{3}(t)\equiv\phi_{4}(t)$
\\
\hline
$S_{1},S_{2}$
&
$0$
&
$\begin{array}{c}
0\\
=:p_{1,0}\end{array}$
&
$\begin{array}{c}
a\\
=:p_{3,0}\approx 0.62307\end{array}$
\\
\hline
$R_{1},R_{2},S'_{3},S'_{4}$&
$\tau$&
$\begin{array}{c}
H_{\epshat}(\tau)\\
=:p_{1,2}\end{array}$&
$\begin{array}{c}
U(p_{3,0}+\tau)+2\epshat>1\\
1\reset0\\
=:p_{3,2}\end{array}$
\\
\hline
$R_{3},R_{4}$&
$2\tau$&
$\begin{array}{c}
H_{2\epshat}(p_{1,2}+\tau)\\
=:p_{1,3}\end{array}$&
$\begin{array}{c}
H_{\epshat}(\tau)\\
=:p_{3,3}\end{array}$
\\
\hline
$S_{1},S_{2}$&
$2\tau+1-p_{1,3}$&
$\begin{array}{c}
1\reset0\\
0=:p_{1,5}\end{array}$&
$\begin{array}{c}
p_{3,3}+1-p_{1,3}\\
=:p_{3,5}\\
\overset{!}{=}a\end{array}$
\\
\hline
\end{tabular}
\end{center}

\caption{
Events occurring during one period of one of the unstable attracting
periodic orbits in the system of four coupled oscillators (see text for
explanation).
\label{Tab:unperturbed_PO}}
\end{table}

Table~\ref{Tab:unperturbed_PO} displays the time evolution on the
periodic orbit event by event. The left column gives the sequence
of events, labelled $S_{i}$ if oscillator $i$ sends a signal and
$R_{i}$ if the signal from oscillator $i$ is received by all other
oscillators $j\neq i$ (for conciseness we write $R_{i}$ here instead
of $\{ R_{i,j}|\, j\neq i\}$). The second left column gives the time
of occurrence of these events. The right two columns give the phases
$\phi_{i}$ of oscillators $i\in\{1,\ldots,4\}$ just  after the events
have occurred, where the synchronized oscillators have identical phases,
$\phi_{1}(t)\equiv\phi_{2}(t)$ and $\phi_{3}(t)\equiv\phi_{4}(t)$.
The first row gives the initial condition at $t=0$, just after the
signals of oscillators $i=1$ and $i=2$ have been sent, $S_{1}$,
$S_{2}$. The initial phases
\begin{equation}
\boldsymbol{\phi}(0)=(0,0,a,a)
\end{equation}
are given
in the first column with constant $a\approx0.62307$ used in
Section~\ref{secuasforpulsecos}, equation (\ref{eq:C22P}) and defined below,
Eq.(\ref{eq:A:C}).

In general, at time $t$ the time interval to the next event is given 
by the minimal distance of any phase to threshold, 
\begin{equation}
\Delta t_1=\min\{1-\phi_{i}(t)\,|\, i\in\{1,\ldots,4\}\},
\end{equation}
or by the time 
\begin{equation}
\Delta t_2=\min\{\tau-\s_{i,k}(t)\,|\, i\in\{1,\ldots,N\}\,
\textrm{and }k\,\textrm{ such that } \s_{i,j}(t)\leq\tau\}
\end{equation}
until the next signal arrives, depending on which $\Delta t_i$ is 
smallest at any given time $t$ (cf.~App.~A).

As an example of the event-based dynamics, consider the event $k=2$
at time $t=\tau$. This event contains the reception of signals sent
by oscillators $i=1$ and $i=2$ at time $t=0$, $R_{1}$ and $R_{2}$.
Since there are no self-connections, these oscillators receive only
one signal each such that their phases are shifted according to
\begin{equation}
p_{i,2}=H_{\epshat}(\tau)\end{equation}
for $i\in\{1,2\}$. At the same time $t=\tau$ the other oscillators
$i=3$ and $i=4$ receive two signals of total strength $2\epshat$
making them superthreshold, $U(p_{i,0}+\tau)+2\epshat>1$ for $i\in\{3,4\}$,
such that they experience a reset, denoted $1\reset0$, and \begin{equation}
p_{3,2}=p_{4,2}=0.\label{eq:A:six0}\end{equation}
Here, one incoming signal would not have been sufficient to reach
the threshold,\begin{equation}
U(p_{i,0}+\tau)+\epshat<1,\end{equation}
for $i\in\{3,4\}$. This ensures that the return map (\ref{eq:F(delta)}) for
perturbations to the periodic orbit is continuous.

Going through the table row by row, the phases proceed (with time)
from event to event. The last row gives the phases just after oscillator
$i=2$ has been reset again,
\begin{equation}
\phi_{2}(T)=p_{2,5}=0,
\end{equation}
after period $T=2\tau-1+p_{1,3}$. The actual parameter dependence
of the constant $a$ is then determined self-consistently from the
condition\begin{eqnarray}
p_{3,5}(a) & \overset{!}{=} & a\end{eqnarray}
for the orbit to be closed. This leads to
\begin{eqnarray}
a & = & H_{\epshat}(\tau)+1-H_{2\epshat}(H_{\epshat}(\tau)+\tau).
\label{eq:A:C}
\end{eqnarray}

\subsection{Dynamics of a general perturbation to the periodic orbit}

\label{Sec:generalperturbation}

Here we consider the dynamics of a general perturbation applied to
the periodic orbit studied above. By general perturbation we mean
that the initial phases
\begin{equation}
\boldsymbol{\phi}(0)=(0,0,a,a)+(\delta_{1},\delta_{2},\delta_{3},
\delta_{4})\label{eq:phi(0)}
\end{equation}
satisfy $\phi_{1}(0)\neq\phi_{2}(0)$. To be specific we assume 
$\delta_{1}>\delta_{2}=0$.
Whether or not $\phi_{3}(0)=\phi_{4}(0)$ does not play a role. Pulses
are then sent by the network in the cyclic order $S_{1},S_{2},(S'_{3},S'_{4})$.
This assumption is made without loss of generality because of the
permutation symmetry; the case $\delta_2>\delta_1=0$ 
can be treated analogously.
We consider only perturbations of firing events that
are locked to the phases such that the firing time variables $\s_{i,k}(t)$ are
unimportant as 
discussed previously.
The changes of the phases due to single events are obtained from Table
\ref{Tab:perturbed_PO}. If we denote $\boldsymbol{\delta}=
(\delta_{1},\delta_{3},\delta_{4})$,
the return map
\begin{equation}
\boldsymbol{\delta}'=\mathbf{F}(\boldsymbol{\delta})\label{eq:F(delta)}
\end{equation}
from the three-dimensional plane defined by $\phi_{2}=0$ to itself
is obtained after all oscillators have fired exactly once (last row
of Table \ref{Tab:perturbed_PO}). %

\begin{table}
{\small
\begin{center}
\begin{sideways}
\begin{tabular}{|l|l|cccc|}
\hline
Event&
$t$&
$\phi_{1}(t)$&
$\phi_{2}(t)$&
$\phi_{3}(t)$&
$\phi_{4}(t)$\\
\hline
$(S_{1}),S_{2}$&
$0$&
$\begin{array}{c}
\delta_{1}\\
=:\p_{1,0}\end{array}$&
$\begin{array}{c}
\delta_{2}=0\\
=:\p_{2,0}\end{array}$&
$\begin{array}{c}
a+\delta_{3}\\
=:\p_{3,0}\end{array}$&
$\begin{array}{c}
a+\delta_{4}\\
=:\p_{4,0}\end{array}$\\
\hline
$R_{1}$&
$\tau-\delta_{1}$&
$\begin{array}{c}
\tau\\
=:\p_{1,1}\end{array}$&
$\begin{array}{c}
H_{\epshat}(\p_{2,0}+\tau-\delta_{1})\\
=:\p_{2,1}\end{array}$&
$\begin{array}{c}
H_{\epshat}(\p_{3,0}+\tau-\delta_{1})\\
=:\p_{3,1}\approx0.9843<1\end{array}$&
$\begin{array}{c}
H_{\epshat}(\p_{4,0}+\tau-\delta_{1})\\
=:\p_{4,1}\approx0.9843<1\end{array}$\\
\hline
$R_{2},S'_{3},S'_{4}$&
$\tau$&
$\begin{array}{c}
H_{\epshat}(\tau+\delta_{1})\\
=:\p_{1,2}\end{array}$&
$\begin{array}{c}
\p_{2,1}+\delta_{1}\\
=:\p_{2,2}\end{array}$&
$\begin{array}{c}
U(\p_{3,1}+\delta_{1})+\epshat>1\,\,\,\\
1\,\reset\,0\\
=:\p_{3,2}\end{array}$&
$\begin{array}{c}
U(\p_{4,1}+\delta_{1})+\epshat>1\,\,\,\\
1\,\reset\,0\\
=:\p_{4,2}\end{array}$\\
\hline
$R_{3},R_{4}$&
$2\tau$&
$\begin{array}{c}
H_{2\epshat}(\p_{1,2}+\tau)\\
=:\p_{1,3}\end{array}$&
$\begin{array}{c}
H_{2\epshat}(\p_{2,2}+\tau)\\
=:\p_{2,3}\end{array}$&
$\begin{array}{c}
H_{\epshat}(\tau)\\
=:\p_{3,3}\end{array}$&
$\begin{array}{c}
H_{\epshat}(\tau)\\
=:\p_{4,3}\end{array}$\\
\hline
$S_{1}$&
$2\tau-1-\p_{1,3}$&
$\begin{array}{c}
1\reset0\\
=:\p_{1,4}\end{array}$&
$\begin{array}{c}
\p_{2,3}+1-\p_{1,3}\\
=:\p_{2,4}\end{array}$&
$\begin{array}{c}
\p_{3,3}+1-\p_{1,3}\\
=:\p_{3,4}\end{array}$&
$\begin{array}{c}
\p_{4,3}+1-\p_{1,3}\\
=:\p_{4,4}\end{array}$\\
\hline
$S_{2}$&
$2\tau+1-\p_{2,3}$&
$\begin{array}{c}
\p_{1,3}-\p_{2,3}\\
=:\p_{1,5}\\
=:\delta_{1}\end{array}$&
$\begin{array}{c}
1\reset0\\
=:\p_{2,5}\\
\,\end{array}$&
$\begin{array}{c}
\p_{3,3}+1-\p_{2,3}\\
=:\p_{3,5}\\
=:a+\delta'_{3}\end{array}$&
$\begin{array}{c}
\p_{4,3}+1-\p_{2,3}\\
=:\p_{4,5}\\
=:a+\delta'_{4}\end{array}$\\
\hline
\end{tabular}
\end{sideways}
\end{center}
}

\caption{Time evolution of a small perturbation to the unstable
periodic orbit as is progresses around one period. A general
perturbation splits up the first synchronized group, 
$\delta_{1}>\delta_{2}\equiv 0$,
while any splitting of the second group is lost due to the event $R_2$, $S'_3$, 
and $S'_4$, i.e. before the first return to the initial section.
This sequence of events will occur for all sufficiently close perturbations to
the orbit. \label{Tab:perturbed_PO} }
\end{table}

Table \ref{Tab:perturbed_PO} displays the time evolution of a general
perturbation event by event. As for the unperturbed orbit described
above, the left column gives the sequence of events and the second
left column gives the time of these events. The right four columns
give the phases of oscillators $i\in\{1,\ldots,4\}$ right after an
event has occurred. The first row gives the initial condition at $t=0$,
just after the signal has been sent, $S_{2}$, by oscillator $i=2$.
Before, at time $t=-\delta_{1}<0$, the signal of oscillator $i=1$
has been sent, denoted $(S_{1})$; this signal will travel until time
$t=\tau-\delta_{1}$. The initial phases (\ref{eq:phi(0)}) are displayed
in the first column. Here $\delta_{2}=0$ by definition. Throughout
the table, the phases after each event are labelled $\varphi_{i,k}$
where the first index labels the oscillator $i$ and the second index
counts the events $k$ starting with $k=0$ for the initial condition as was also
used in Table~\ref{Tab:unperturbed_PO} above.

For instance, the first event (after the initial condition), $k=1$,
at time $t=\tau-\delta_{1}$ is the reception $R_{1}$ of the signal
sent by oscillator $i=1$. Since there are no self-interactions, the
phase of oscillator $i=1$ is only shifted in time,
\begin{equation}
\varphi_{1,1}=\delta_{1}+(\tau-\delta_{1})=\tau.
\end{equation}
The phases of all other oscillators $i\in\{2,3,4\}$ additionally
jump because of the incoming subthreshold signal,
\begin{equation}
\varphi_{i,1}=H_{\epshat}(\varphi_{i,0}+(\tau-\delta_{1})).
\end{equation}

Again, as for the unperturbed dynamics, going through the table row
by row, the phases are calculated event by event. The phases $\varphi_{i,k}$
for $i\in\{1,\ldots,4\}$ right after the events are approximated
by the unperturbed phases $p_{i,k}$ defined in Table \ref{Tab:unperturbed_PO},
\begin{equation}
\varphi_{i,k}=p_{i,k}+\mathcal{O}(\boldsymbol{\delta})
\end{equation}
for all $k$ for which the $p_{i,k}$ are defined, $k\in\{0,2,3,5\}$.
Here $\mathcal{O}(\boldsymbol{\delta})$ denotes a function of
all $\delta_{i}$, $i\in\{1,3,4\}$, that approaches zero at
least linearly if all $\delta_{i}\rightarrow0$, i.e.\
$\mathcal{O}(\boldsymbol{\delta}):=\sum_{i=1}^{N}\mathcal{O}(\delta_{i})$.

We  thus obtain the return map\begin{eqnarray}
\mathbf{F}(\boldsymbol{\delta}) & = & \boldsymbol{\delta}'\\
 & = & (\delta'_{1},\delta'_{3},\delta'_{4})\\
 & = & (\varphi_{1,5},\varphi_{3,5}-a,\varphi_{4,5}-a)\end{eqnarray}
from the last row of Table \ref{Tab:perturbed_PO}. Here the $\varphi_{i,5}$
again depend on the original perturbation vector $\boldsymbol{\delta}$.
From this identification we obtain the components
\begin{equation} \left.
\begin{array}{lll}
F_{1}(\boldsymbol{\delta}) & = & H_{2\epshat}(H_{\epshat}(\tau+\delta_{1})+\tau)
-H_{2\epshat}(H_{\epshat}(\tau-\delta_{1})+\tau+\delta_{1})\\
F_{3}(\boldsymbol{\delta})=F_4(\boldsymbol{\delta}) & = &
H_{2\epshat}(H_{\epshat}(\tau)+\tau)\quad\quad\,-H_{2\epshat}
(H_{\epshat}(\tau-\delta_{1})+\tau+\delta_{1})
\end{array}\right.\label{eq:A:F_i}\end{equation}
Note that the component functions $F_{i}(\boldsymbol{\delta})$ in
(\ref{eq:A:F_i}) are of order
$F_{i}(\boldsymbol{\delta})=\mathcal{O}(\boldsymbol{\delta})$;
moreover the dynamics clearly depends only on the initial
 perturbation $\delta_1$ to this orbit.
It is important to note that although $\delta_{3}\neq0$ and $\delta_{4}\neq0$
in general, this group is resynchronized at the next return,
$\delta'_{3}=\delta'_{4}$.

\end{document}